\documentclass[a4paper,11pt]{article}
\pdfoutput=1 

\usepackage{jcappub} 

\usepackage[T1]{fontenc} 
\usepackage{subfigure}
\usepackage{color}%

\title{\bf{Dynamical systems analysis of an interacting scalar field model in an anisotropic universe}}

\author[a]{Sujoy Bhanja,}
\author[a]{Goutam Mandal,}
\author[b]{Abdulla Al Mamon}
\author[a]{and Sujay Kr. Biswas}

\affiliation[a]{ Department of Mathematics, University of North Bengal, Raja Rammohunpur, Darjeeling-734013, West Bengal, India.}
\affiliation[b]{Department of Physics, Vivekananda Satavarshiki Mahavidyalaya (affiliated to the Vidyasagar University), Manikpara 721513, West Bengal, India.}

\emailAdd{sujoybhanja@gmail.com; 
	rs\_sujoy@nbu.ac.in}
\emailAdd{gmandal243@gmail.com; 
	rs\_goutamm@nbu.ac.in}
\emailAdd{abdulla.physics@gmail.com}
\emailAdd{sujaymathju@gmail.com; sujay.math@nbu.ac.in}

\abstract{In this paper, we investigate a non-canonical scalar field model in the background dynamics of anisotropic Locally Rotationally Symmetric (LRS) Bianchi type I universe where gravity is coupled minimally to scalar field which is taken as dark energy and pressureless dust as dark matter are the main matter content of the universe. We perform dynamical system analysis to characterize the cosmological evolution of the model with and without interaction in the dark sector separately. First, we convert the evolution equation into an autonomous system of ordinary differential equations by using a suitable choice of dimensionless variables, which are normalized over the Hubble scale.  We choose scalar field coupling and potential in such a way that the autonomous system converted to a 2D system. Linear stability theory is employed to the extracted critical points to find the nature. From the analysis, we find some  interesting cosmological scenarios, such as late-time scalar-field dominated solutions, which evolve in the quintessence era, cannot solve the coincidence problem. Accelerated scaling attractors are also obtained that correspond to the late phase evolution in agreement with present observational data, and these solutions also provide possible mechanisms to alleviate the coincidence problem. A complete cosmic evolution is obtained from early inflation to a late-time dark energy-dominated phase, connecting through a matter-dominated transient phase of the universe. Furthermore, we find that for different values of the interaction parameter $\alpha$, the evolutionary trajectories of the Hubble parameter, and the distance modulus forecasted by the model are in quite well agreement with observational datasets.}

\arxivnumber{1}

\begin{document}
\maketitle
\flushbottom

\section{Introduction}
According to several cosmological observations\cite{Riess:1998cb, Perlmutter:1998np, P. Ade,Scranton, Tegmark,Spergel}, the presently accelerated expansion phase of the universe has been well established. In general relativity, dark energy (DE), a mysterious component of energy, is supposed to be responsible for this observed acceleration mechanism, and one can look into Refs. \cite{cst2006,bamba2012} for a detailed review of DE. However, the origin and nature of DE are completely unknown to us except for its large negative pressure. The most simple cosmological model for DE is the $\Lambda$CDM model, which is very well consistent with recent observations. This model consists of a cosmological constant  parameter $\Lambda$  along with a cold dark matter (CDM) component. The energy density of the cosmological constant remains constant with time and $\Lambda$ is characterized by equation of state (EoS) parameter $\omega_{\Lambda}=-1$. Although the $\Lambda$CDM model has many successes, the cosmological constant  parameter suffers from the  {\it coincidence} and {\it fine-tuning}  problems (Why the value of $\Lambda$ is very minimal when derived
in natural units and since it could have two different contributions from the matter part and the gravitational part, why does the sum remain so fine tuned $?$) \cite{Weinberg,T. Padmanabhan2003,Steinhardt}. In this context, some alternative theories have been proposed for the dynamical DE scenario to resolve these issues (for a review, one can see Ref. \cite{labookcup} and the references therein).\\ 

\par The simple and popular dynamical DE model is the quintessence (canonical scalar field) \cite{Copeland, Caldwell,Chimento,Samanta,Chaubey}, whose EoS parameter is in the range of $\omega_{\phi}\in (-1,-\frac{1}{3})$. On the other hand, the phantom DE model \cite{Roy and Bhadra, Ludwick, Mahata,R.R.Caldwell,Robert J. Scherrer,Anjan A Sen} is described by the EoS parameter  $\omega_{\phi}<-1$, which according to observational data, cannot be ruled out, and moreover, it has a great interest in cosmological study as the cosmological dynamics of this model has valuable contributions to describe the current acceleration of the universe. It is worthwhile to mention here that unified model as a unification of the DE and dark matter (DM) is  another possible candidate among various DE models. For a discussion on the unified DE model, we refer the reader to Refs. \cite{Kamenshchik,Bertacca, Sujay Kr. Biswas and  Atreyee Biswas, Abdulla Al Mamon} and the references therein. Generally, in a scalar field model when a coupling function (depending on scalar field $\phi$) as a non-canonical form  appears with a kinetic term as a pre-factor in both the energy density and pressure terms, the scalar field model is called the non-canonical scalar field \cite{Bertacca,Goutam Mandal,Nicola Tamanini}. In particular, the phantom scalar field having a non-canonical negative kinetic term is reverse to the quintessence scalar field model giving a cosmological scenario with $\omega_\phi<-1$ in its evolution \cite{P. Ade}. Also, in this  model, a scalar field is minimally coupled to gravity to drive the accelerated expansion phase of the Universe. Indeed, various interesting possibilities with non-canonical scalar field has been recently studied in the literature for addressing several kinds of cosmological issues \cite{nsfm11,nsfm21,nsfm31,nsfm41,nsfm51,nsfm61,nsfm71,nsfm81,nsfm91}. \\

\par As is well-known, the dark sectors (DE and DM) are the dominant source of the total energy budget of the universe, so one cannot neglect to consider an interaction between these components. Naturally, an interacting scenario, which allows exchanging the energies between dark sectors, can give a better dynamics than that of a non-interacting one. In this context, different interacting DE models have been widely studied in literature \cite{Goutam Mandal,Liu,G. Leon,P. Silveira,R. Lazkoz,R. C. Nunes,S. D. Odintsov,T. Gonzalez}. Some recent studies indicate  that the interaction can give a possible mechanism to alleviate the $H_0$ tension as well \cite{P. D. Alvarez,J. P. Hu}. \\ 

\par According to the cosmological principle, the universe is assumed to be homogeneous and isotropic at large scale and the Fiedmann-Lemaitre-Robertson-Walker (FLRW) Universe describes it very well. However, this is challenged by some recent puzzling cosmological observations \cite{Netterfield,D.N. Spergel}, and such cosmic anomalies exhibit the anisotropy nature of the universe.
Also the experimental data indicates the anisotropy in the universe's evolution that directs the existence of anomalies in the CMB (Cosmic Microwave Background), suggesting the plane mirroring symmetry when we assume large scale \cite{Gurzadyan,Gurzadyan1}. Therefore, a slight deviation from isotropy to anisotropy to the CMB is not of little significance.  There will be a small anisotropy with a small ($10^{-5}$) amplitude. Based on observational evidence, other types of deviations from isotropy are potentially connected to the large-angle anomalies in the CMB \cite{Bennet}. Among these anomalies, some remarkable possibilities, such as "the large-scale symmetry \cite{Eriksen2004,Eriksen2004a,Hansen}" and "the low quadrupole moment of the CMB to a Bianchi type I anisotropic evolution of the universe
\cite{Barrow,Berera,Campanelli2006,Campanelli2006a,Campanelli12007}", in which an ellipsoidal expansion of the universe is indicated because at large scale the low quadrupole moment of CMB is suppressed.
Hence, the observational data from various sources give birth to a new question on the homogeneity and isotropy of the standard cosmological model and simultaneously indicate the anisotropic behavior of the universe, atleast at local-scale \cite{Pranjal}. As a result, some strange motion exists in the observable universe because of the local inhomogeneity and anisotropy of the surrounding structures, which cannot be neglected \cite{J. Colin}. Anisotropic characteristics of the universe can be explained by describing the universe's large- scale behavior where spatially homogeneous and anisotropic cosmological models play a vital role. These cosmological models have been extensively studied in \cite{D. C. Maurya,Yadav L,Kumar S,Pradhan A,Adhav KS,Adhav,Raushan,T. Singh,Sobhanbabu} to achieve the picture of an early stage of the universe . Naturally, one simple and convincing cosmological model is required to allow directional scale factors and  maintain the homogeneity and flatness. The Bianchi type I is the model which fulfills the above criteria. Some anisotropic cosmological models have been studied widely in the literature \cite{D. C. Maurya,Yadav L,Kumar S,Pradhan A,Adhav KS,Adhav,Raushan,T. Singh,Sobhanbabu,Misner1973,Peebels1993,L. Tedesco,Suresh Kumar,L. Perivolaropoulos,D. Paul,Bijan Saha,Rishi,Paliathnasis2022,Fadragas2014,Escobar2012}. Interestingly, Bianchi type I provides a minimal deviation from isotropy, so it can be considered an excellent alternative to the standard FLRW model to study the various cosmic anomalies in the CMB \cite{Campanelli2011,Wang,Luigi}.  Hence, it is appropriate to study the anisotropic natutre of universe's evolution in its early and present phase.\\


In an anisotropic universe, the Bianchi space-times are generally described by three scale factors. The Locally Rotationally Symmetric (LRS) Biachi metric has two independent scale factors due to its extra isometry. Now, it is interesting to mention that these geometries give engaging physical scenarios of inflation as well as post-inflation epochs. The primary motivation of this work is to undertake the LRS Bianchi Type I universe as a background dynamics in which an interacting scalar field model is investigated to achieve an overall evolution of the universe from early inflation to late time dynamics.
In particular, we investigate the non-canonical scalar field model minimally coupled to gravity in the background dynamics of spatially flat, homogeneous, and Bianchi type I anisotropic model of the universe. Here, a pre-factor $\lambda(\phi)$ in the kinetic term  is considered as a coupling function. As the evolution equation is very complicated in form so, dynamical system analysis \cite{Wainwright,Mandal1,Coley} is performed to study the feature of evolution qualitatively.
We consider here a Locally Rotationally Symmetric (LRS) metric in Bianchi I model where scalar field is taken as DE, and pressureless dust as DM. An interaction term is also considered in the dark sector. A pre-factor (coupling function) is assumed to be a function of a real scalar field $\phi$ appeared in the kinetic term of scalar field. We adopt dynamical systems tools to extract overall information of evolution from the model, bypassing the non-linearities and complications of governing equations. We consider two dimensionless variables normalized with respect to the Hubble parameter.
We choose the coupling function and potential of scalar field in such a way that the autonomous system becomes two-dimensional. Critical points from two-dimensional autonomous systems exhibit cosmological interested phenomena such as from early inflation to late time cosmology. From the dynamical analysis, we obtain some physically relevant scenarios. Some critical points show the late-time evolution of the universe dominated by quintessence and evolved in quintessence and cannot solve the coincidence problem. Some of the points predict the late-time scaling attractors supported by present observational data and solve the coincidence problem. Some points refer to the complete evolutionary scheme of the universe, starting from early inflation to late-time dark energy-dominated acceleration via a matter-dominated intermediate phase. \\

\par The paper is organized as follows: In the next section, we analyze the dynamics of a non-canonical scalar field which is coupled to gravity with a coupling parameter in the background of LRS Bianchi type I metric. Section \ref{sec-dyna} comprises the formation of an autonomous system and the physical parameters. Phase space analysis has also been performed. Cosmological implications of the critical points are discussed in  section \ref{sec-cp}. Furthermore, we compared the theoretical model with the observational data in section \ref{cdata}. Lastly, section \ref{sec-last} is allocated to some concluding remarks. Throughout the paper, we use the natural units $k^{2}=8\pi G=c=1$ and the metric convention as (+, - ,- ,-).

\section{Basic equations in anisotropic model}\label{model and autonomous system}
We consider the background dynamics of anisotropic Bianchi type I universe where  scalar field as DE is minimally coupled to the gravity. The scalar field in presence of pre-factor (coupling term) with kinetic term is called non-canonical. The action of this non-canonical scalar field  Lagrangian takes the following form\cite{Mahata,J. Dutta,S. Mishra} 
\begin{equation}\label{action}
	S=\int d^4x\sqrt{-g}\Bigg[\frac{R}{2k^{2}}-\frac{1}{2}\lambda(\phi)g^{\mu\nu}(\nabla_\mu \phi)(\nabla_\nu \phi)-V(\phi)\Bigg]+S_M.
\end{equation}
where  $\lambda(\phi)$, an arbitrary function of scalar field $\phi$, is the coupling in scalar field
and  $V(\phi)$ is the potential of the scalar field. The gravitational coupling is $\kappa^{2}=8\pi G$, where $G$ is the Newton's gravitational constant. The action $S_M$ denotes the action for the matter content.
We consider particularly the locally rotationally symmetric (LRS) Bianchi type I metric given in the following
\begin{equation}\label{metric}
	ds^{2}=dt^{2}-A(t)^{2}dx^{2}-B(t)^{2}\left(dy^{2}+dz^{2}\right),
\end{equation}
where $A(t)$ and $B(t)$ are the directional scale factors depended on cosmic time $t$. The scale factor $A(t)$ measures the expansion in $x$-direction while $B(t)$ scales in the $y$ and $z$-directions respectively. Hubble parameters in different axial directions are denoted by  $H_1$ and $H_2$ and are defined by
\begin{equation}\label{individual Hubble}
	H_{1}=\frac{\dot{A}}{A}, ~~~~ H_{2}=\frac{\dot{B}}{B},
\end{equation}
where over-dot denotes differentiation with respect to cosmic time $t$. It is obvious that the usual FLRW metric is recovered for $A=B$. We now introduce an average scale factor $a(t)$ in this LRS Bianchi type I metric as
\begin{equation}\label{scale factor}
	a(t)=\left(AB^{2} \right)^{\frac{1}{3}},
\end{equation}
Therefore, the average Hubble parameter $H$ is given by
\begin{equation}\label{average Hubble}
	H=\frac{\dot{a}}{a}  =\frac{1}{3}\left(\frac{\dot{A}}{A}+2\frac{\dot{B}}{B} \right)=\frac{1}{3}\left(H_1+2H_2 \right).
\end{equation}
The energy conservation equation for the matter takes the form
\begin{equation}\label{conservation ind Hubble}
	\dot\rho+\left(\frac{\dot{A}}{A}+2\frac{\dot{B}}{B} \right)\left(\rho+p \right)=0,
\end{equation}
which reduces to the following form in terms of average Hubble parameter $H$
\begin{equation}\label{conservation tot}
	\dot\rho+3H \left(\rho+p \right)=0,
\end{equation}
It should be mentioned that in this work, we neglect the anisotropies of the matter content of the universe, rather we focus only on the geometrical anisotropies. Therefore, we consider here that the matter content, namely matter and scalar field to be homogeneous and isotropic in nature. The volume expansion is defined by the average Hubble parameter and directional Hubble parameters as follows
\begin{equation}\label{expansion}
	\Theta=3H=\left(H_1+2H_2 \right).
\end{equation}
The shear scalar due to anisotropy of geometry of the space-time is given by 
\begin{equation}\label{shear scalar}
	\sigma^{2}=\frac{1}{2}\sigma_{\mu\nu}\sigma^{\mu\nu},
\end{equation}
where $\sigma_{\mu\nu}$ is the shear tensor \cite{B. C. Paul}.
In the LRS Bianchi-I anisotropic space-time given by Eqn. (\ref{metric}), the shear tensor expresses in following way
\begin{equation}\label{shear tensor}
	\sigma_{\mu\nu}\sigma^{\mu\nu}=\sum_{i=1}^{3}\left(H_i-H\right)^{2}=H_1^{2}+2H_2^{2}-3H^{2}.
\end{equation}
For this model, the average anisotropic expansion rate ($\it{A_n}$) is given by
\begin{equation}\label{anisotropic expansion}
	\it{A_n}=\frac{1}{3}\sum_{i=1}^{3}\left(\frac{\Delta H_i}{H} \right)^{2},
\end{equation}
where $\Delta H_i=H_i-H$. 
Now, by using Eqns. (\ref{average Hubble}) and (\ref{shear tensor}), the shear scalar ($\sigma^{2}$) (in Eqn.  (\ref{shear scalar})) takes the form
\begin{equation}\label{shear scalar Hubble}
	\sigma^{2}=\frac{1}{3}\left(H_1-H_2 \right)^{2}.
\end{equation}
Here, we assume the anisotropic scenario in which the shear $(\sigma)$ and expansion scalar $(\Theta)$ follow the proportional relation $\sigma\propto \Theta$ (see in Ref. \cite{Collins,Rao}). As a result, for constant $\frac{\sigma}{\Theta}$, one obtains $\frac{\dot{A}}{A}=n\frac{\dot{B}}{B}$ which implies that $A=kB^n$, where $k$ is an integrating constant and $n$ is any positive constant.  Without any loss of generality one can take the integrating constant $k$ as unity (i.e., $k=1$). Note that for $n=1$, the model recovers the usual FLRW metric and for any other values of $n$ the model becomes anisotropic. From the above we obtain the following relation:
\begin{equation}\label{A B relation}
	A\propto B^{n},
\end{equation}
Now from the expressions (\ref{individual Hubble}), (\ref{average Hubble}) and (\ref{A B relation}), we obtain the following relations of directional Hubble parameters and average Hubble parameter through the constant $n$ (by taking $k=1$):
\begin{equation}\label{Hubble relns}
	H_1=nH_2=\frac{3n}{n+2}H.
\end{equation}
The shear scalar in Eqn.(\ref{shear scalar Hubble}) will take the following form
\begin{equation}\label{shear scalar n}
	\sigma^{2}=3\left(\frac{n-1}{n+2} \right)^{2}H^{2}.
\end{equation}
In the spatially flat, homogeneous, LRS Bianchi I universe the modified  Einstein's equations become
\begin{equation}\label{Friedmann dircn}
	2\frac{\dot{A}\dot{B}}{AB}+\left(\dfrac{\dot{B}}{B} \right)^{2}=\left(\rho_{M}+\rho_{\phi} \right),
\end{equation}
\begin{equation}\label{accl dircn1}
	\dfrac{\ddot{A}}{A}+\frac{\ddot{B}}{B}+\frac{	\dot{A}\dot{B}}{AB}=-p_\phi,
\end{equation}
\begin{equation}\label{accl dircn2}
	2\frac{\ddot{B}}{B}+\left(\frac{\dot{B}}{B} \right)^{2}=-p_\phi.
\end{equation}
Where $\rho_{\phi}$ and $p_\phi$ represent the energy density and pressure of the scalar field ($\phi$) respectively and $\rho_{M}$ denotes the energy density of the matter. For this scalar field taken in action (\ref{action}), the thermodynamic pressure and energy density are as
\begin{equation}\label{scalarfield pressure}
	p_\phi=\frac{1}{2}\lambda(\phi)\dot{\phi}^{2}-V(\phi)
\end{equation}
and 
\begin{equation}\label{scalarfield energy}
	\rho_{\phi}=\frac{1}{2}\lambda(\phi)\dot{\phi}^{2}+V(\phi).
\end{equation}
We consider the main contents of the universe are scalar field as DE and pressure-less dust ($p_M=0$) as DM. They are the dominant source of the universe. In presence of an interaction term $Q$, the individual energy conservation equations for two dark sectors (DE and DM) are
\begin{equation}\label{conservation DM}
	\dot\rho_{M}+3H\rho_{M}=Q,
\end{equation}
and 

\begin{equation}\label{conservation DE}
	\dot\rho_{\phi}+3H\left(\rho_{\phi}+p_\phi \right)=-Q
\end{equation}
respectively. Positivity of $Q$ indicates the occurrence energy exchange in the direction of DE from DM and it happens in the reverse direction for the negativity of $Q$.
From the equations (\ref{scalarfield pressure}), (\ref{scalarfield energy}) and (\ref{conservation DE}), we get evolution equation for scalar field with coupling function:
\begin{equation}\label{evolution}
	\lambda(\phi)\ddot{\phi}+\lambda' \frac{\dot{\phi}^{2}}{2}+3H\lambda(\phi)\dot{\phi}+V'(\phi)~ = ~ -\frac{Q}{\dot{\phi}}.
\end{equation}
We now obtain the modified Friedmann constraint and the acceleration equation from the equations (\ref{Friedmann dircn}), (\ref{accl dircn1}) and (\ref{accl dircn2}) by using equations (\ref{Hubble relns}), (\ref{scalarfield pressure}) and (\ref{scalarfield energy}). Therefore, the modified Friedmann equation and the acceleration equation will take the following form in terms of average Hubble parameter as
\begin{equation}\label{Friedmann}
	H^{2}=\dfrac{(n+2)^{2}}{9(2n+1)}\left(\rho_{M}+\dfrac{1}{2}\lambda(\phi)\dot{\phi}^{2}+V(\phi)\right)
\end{equation} 
and
\begin{equation}\label{acceleration}
	\dot{H}=-\dfrac{(n+2)^{2}}{6(2n+1)}\left(\rho_{M}+\lambda(\phi)\dot{\phi}^{2}\right).
\end{equation} 
Note that the Friedmann equation (\ref{Friedmann dircn}) can be rearranged in terms of shear scalar ($\sigma^2$):
\begin{equation}\label{Friedman shear}
	H^{2}=\dfrac{(n+2)^{2}}{9n(4-n)}\left(\rho_{M}+\rho_{\phi}-3 \sigma^{2}\right).
\end{equation}
The shear scalar in Eqn. (\ref{shear scalar n}) leads to the anisotropic density parameter $(\Omega_\sigma)$ in terms of anisotropic expansion parameter as
\begin{equation}\label{density anisotropic}
	\Omega_\sigma=\frac{\sigma^{2}}{3H^{2}}=\left(\frac{n-1}{n+2} \right)^{2}.
\end{equation}
where $n>0$. It measures the anisotropies in the space-time dynamics of the model considered here. Note that $n=1\Longrightarrow \Omega_\sigma=0$ and $A_n$ (measures the deviation from anisotropic expansion) vanishes at $n=1$ and this indicates that the model becomes the FLRW universe.  



\section{Dynamical analysis of the model}\label{sec-dyna}

As the evolution  equations are non-linear and complicated in form, exact analytical solutions cannot be obtained. Therefore, we shall apply dynamical system tools and techniques to extract the informations of evolution from the model because this analysis allows us to bypass the non-linearities and complications of cosmological evolution equations. For that one has to convert the evolution equations into an autonomous system of Ordinary Differential Equations (ODEs) by proper transformation of variables. We consider below the dynamical variables in terms of cosmological variables
\begin{equation}\label{variables}
	x^{2}=\frac{(n+2)^{2}}{9(2n+1)}\frac{\lambda(\phi)\dot{\phi}^{2}}{2H^{2}},~~~~~~~~~~~ y^{2}=\dfrac{(n+2)^{2}}{9(2n+1)}\dfrac{V(\phi)}{H^{2}}
\end{equation} 
The dimensionless variables are normalized over Hubble scale.
In this regard we introduce a new independent variable, called e-folding parameter
\begin{equation}\label{e folding}
	N=\ln a= \frac{1}{3} \ln \left(A B^{2} \right).
\end{equation}
As we are concerned with the asymptotic behavior of the evolution, so we take derivatives of x and y with respect to the number of e-folding N as independent variable. After some calculations, the governing equations lead to the following system of ordinary differential equations in terms of dynamical variables
\begin{eqnarray}\label{non-autonomous}
	\begin{split}
		\frac{dx}{dN}& =\frac{3}{2} x \left(x^2-y^2-1\right)-\frac{\beta  \left(\sqrt{3} \sqrt{2 n+1}\right) y^2}{n+2}-\frac{Q x}{\lambda(\phi) \dot{\phi}^{2} H},& \\
		\frac{dy}{dN}& =y \left(\frac{\beta  \left(\sqrt{3} \sqrt{2 n+1}\right) x}{n+2}+\frac{3}{2} \left(x^2-y^2+1\right)\right),
		&~~
	\end{split}
\end{eqnarray}
where $\beta=\frac{\sqrt{3} V'(\phi)}{\sqrt{2}V(\phi) \sqrt{\lambda(\phi)}}$ is a function of scalar field $\phi$. Note that the potential function $V(\phi)$ and coupling function $\lambda(\phi)$ can be chosen independently. In a particular case, one may consider $V(\phi)$ and $\lambda(\phi)$ as exponential, power-law or any other form. In fact, we shall consider $V(\phi)$ and $\lambda(\phi)$ in such a way that the term $\beta$ will become a constant so as to make the autonomous system a two- dimensional and it is easy to analyse mathematically.  For that cases, the choices can be made in the following ways \cite{Mahata}:
\begin{enumerate}
	\item $ V=V_0 ~\mbox{exp}~ (\mu \phi) $, ~~~$\lambda(\phi)=\mbox{constant}$
	
	\item $V=V_0~ \mbox{exp} \big[ \sqrt{\frac{2\lambda_0}{3}} \frac{\xi}{d+1} \phi^ {d+1}\big]$, ~~~~$\lambda({\phi}) =\lambda_{0} \phi^{2d}  $
	\item $V=V_0 \phi^{\sqrt{\frac{2\lambda_{0}}{3}} \xi}$, ~~~~$\lambda(\phi)=\lambda_{0}\phi^{-2}$
	
\end{enumerate}
The above choices are extracted through the relation:
$V(\phi)=V_0 \mbox{exp} \big[\sqrt{\frac{2}{3} }~\xi \int \sqrt{\lambda(\phi)} d\phi \big]$.	

Using the dynamical variables considered in Eqn. (\ref{variables}), we obtain the physical parameters in terms of dynamical variables. The density parameter for DM takes the form 
\begin{equation}\label{density DM}
	1-x^{2}-y^{2}=\dfrac{\rho_M}{3m^{2}H^{2}}=\Omega_M
\end{equation}
The density parameter for the scalar field $\phi$ has the form
\begin{equation}\label{density DE}
	\Omega_\phi=\frac{\rho_\phi}{3m^{2} H^{2}}=x^{2}+y^{2},
\end{equation}
the equation of state parameter for the scalar field is given by 
\begin{equation}\label{EoS DE}
	w_\phi=\frac{p_\phi}{\rho_\phi}=\dfrac{\frac{1}{2}\lambda(\phi)\dot{\phi}^{2}-V(\phi)}{\frac{1}{2}\lambda(\phi)\dot{\phi}^{2}+V(\phi)}=\frac{x^{2}-y^{2}}{x^{2}+y^{2}},
\end{equation}
and the expression of effective equation of state parameter has the form
\begin{equation}\label{effective EoS}
	w_{eff} =\frac{p_M+p_\phi}{\rho_M+\rho_\phi}=\dfrac{\frac{1}{2}\lambda(\phi)\dot{\phi}^{2}-V(\phi)}{\rho_M+\frac{1}{2}\lambda(\phi)\dot{\phi}^{2}+V(\phi)}=x^{2}-y^{2}.
\end{equation}
The deceleration parameter reads as
\begin{equation}\label{deceleration}
	q=-1-\dfrac{\dot{H}}{H^{2}}=\frac{1}{2}\left(1+3x^{2}-3y^{2}\right)=\frac{1}{2}\left(1+3w_{eff}\right),
\end{equation}
which indicates the condition for acceleration when $q<0$, i.e.,  $w_{eff}<-\frac{1}{3}$. For this model, the constraint equation 
(\ref{Friedmann}) gives the feasible region in the phase plane. The energy condition $ 0\leq \Omega_{M} \leq 1$ gives restriction  in physical region of dynamical variables as
\begin{equation}\label{region}
	0\leq x^2+y^2 \leq 1
\end{equation}
We shall undertake a new parameter $m= \frac{  \sqrt{3} \sqrt{2 n+1} }{n+2}$ throughout the work for mathematical simplicity only. For $n>0$, one obtains the restrictions in $m$ as $-1\leq m<0,~~ 0<m\leq1$. We rewrite the anisotropic density parameter ($\Omega_\sigma$) (in Eqn. (\ref{density anisotropic}) ) 
in terms of $m$ as:
\begin{equation}\label{density anisotropic m}
	\Omega_\sigma=1-m^{2}.
\end{equation}
If $m=\pm 1$ then $\Omega_\sigma=0$ which indicates that the model recovers FLRW model.

\subsection{Non-interacting case $Q=0$}\label{non-interacting}

In this subsection we shall study phase space analysis of the LRS Bianchi I model of universe for uncoupled case only i.e., the model without interaction term. Using $Q=0$ in the system of ODEs in Eqn.(\ref{non-autonomous}) we get the following autonomous system:
\begin{eqnarray}\label{autonomous noninteracting}
	\begin{split}
		\frac{dx}{dN}& =\frac{3}{2} x \left(x^2-y^2-1\right)-\beta  m y^{2},& \\
		\frac{dy}{dN}& =y \left(\beta m x+\frac{3}{2} \left(x^2-y^2+1\right)\right),
		&
	\end{split}
\end{eqnarray}
where $\beta$ is constant parameter, takes any real value and $m$ satisfies $-1\leq m<0, ~~0<m \leq 1$. 

\subsubsection{Critical points and phase plane analysis} 
\begin{figure}
\centering
\subfigure[]{%
	\includegraphics[width=10cm,height=10cm]{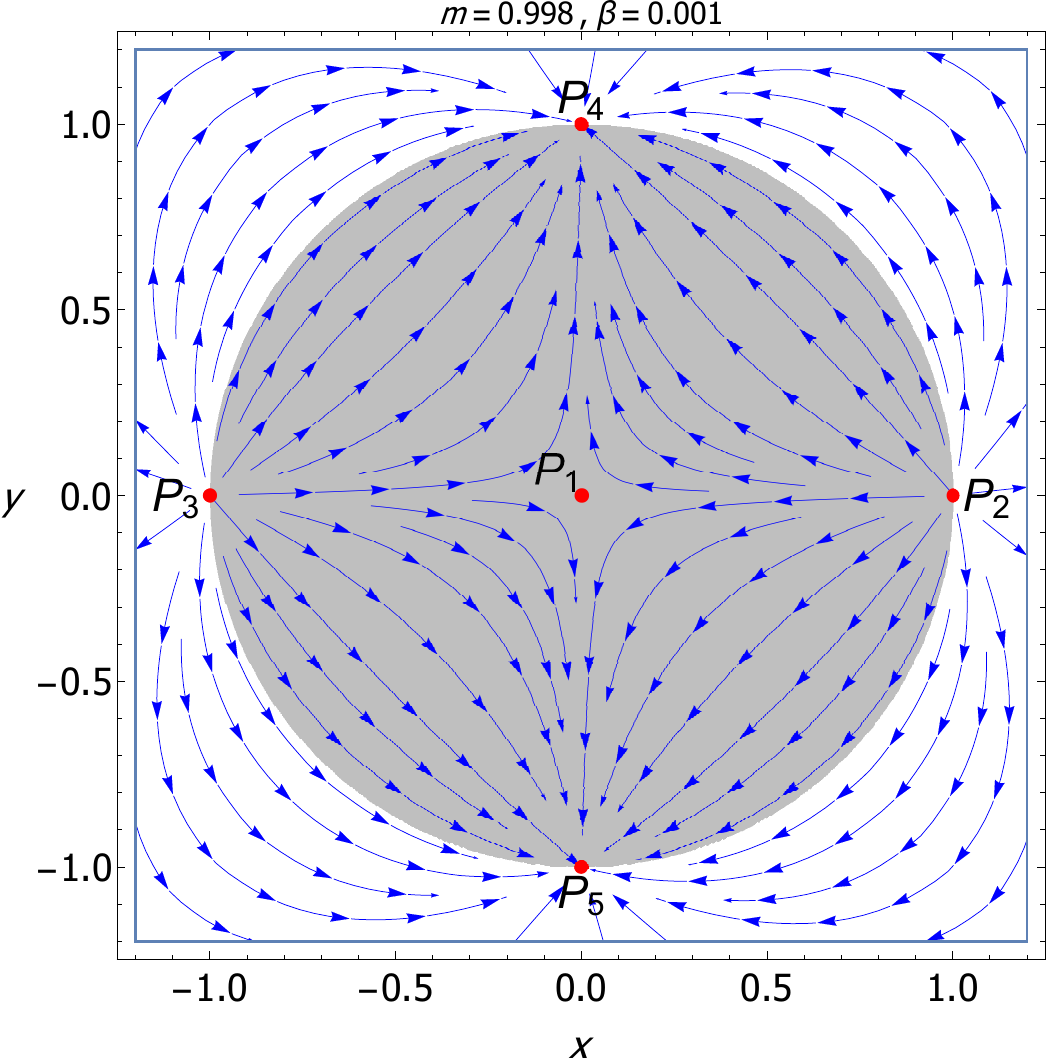}\label{noninteracting}}
\qquad
\subfigure[]{%
	\includegraphics[width=12cm,height=8cm]{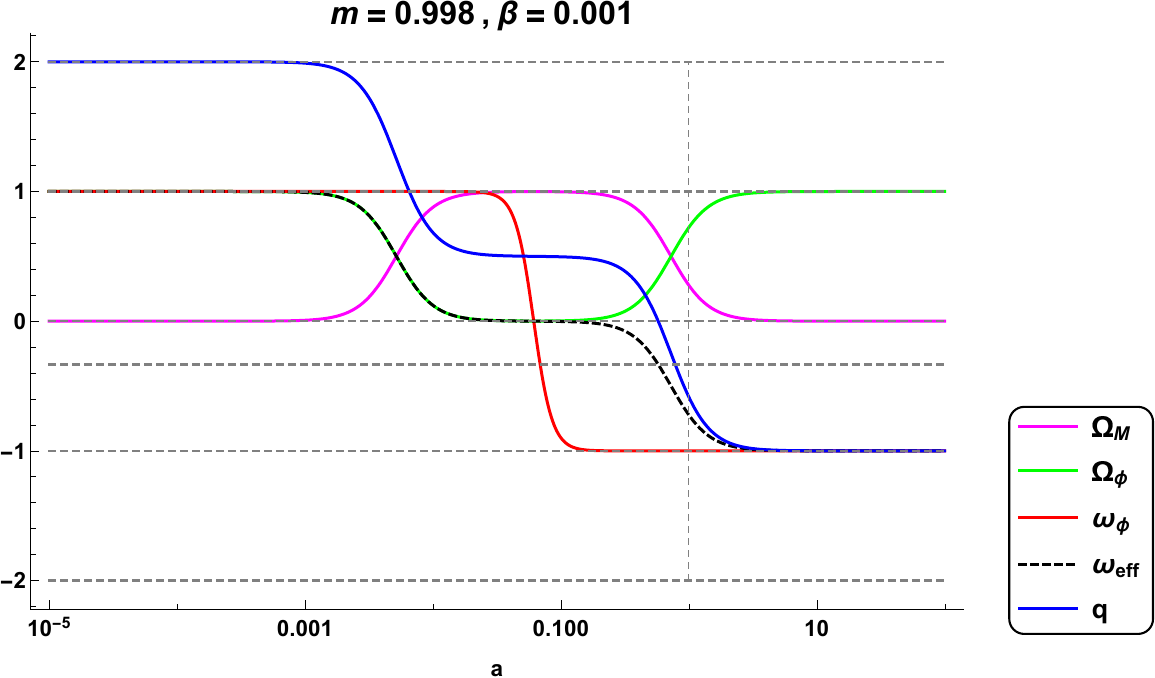}\label{noninteracting evolution}}

\caption{The figure shows the evolution of cosmological parameters and the the phase plane of the autonomous system (\ref{autonomous noninteracting}) for the parameters values $m=0.998$ and $\beta=0.001$. Panel (a) shows that matter dominated $P_1$ is saddle solution, $P_4$ and $P_5$ scalar field dominated late time attractors and the points $P_2$ and $P_3$ are source but decelerated. Panel (b) shows the evolution of cosmological parameters where the late time evolution of the universe attracted by cosmological constant connecting through a matter dominated decelerated phase.}
\label{uncoupled}
\end{figure}

In this section, we shall find out the critical points from $2D$ autonomous systems and also we shall carry out the local stability of the present anisotropic model.
The critical points for the autonomous systems (\ref{autonomous noninteracting}) are the following
{\bf 
	\begin{itemize}
		\item Critical Point $P_1:~(0,~0)$
		\item Critical Point $P_2:~(1,~0)$
		
		\item Citical Point $P_3:~(-1,~0)$
		
		\item Critical Point $P_4:~\left(-\frac{\beta m }{3},~\sqrt{1-\frac{\beta^{2}m^{2}}{9}}\right)$
		
		\item Critical Point $P_5:~\left(-\frac{\beta m }{3},~-\sqrt{1-\frac{\beta^{2}m^{2}}{9}}\right)$
		\item Critical Point $P_6:~\left( -\frac{3}{2\beta m} ,~\frac{3}{2\beta m} \right)$
		\item Critical Point $P_7:~\left( -\frac{3}{2\beta m} ,~-\frac{3}{2\beta m} \right)$	
	\end{itemize}
	
}
\begin{table}[h!] \centering
	\caption{The critical points and the relevant physical parameters  are presented.}%
	\begin{tabular}
		[c]{cccccccc}\hline\hline
		\textbf{Critical Point}&$\mathbf{\Omega_{M}}$& $\mathbf{\Omega_{\phi}}$ & $\mathbf{\omega_{\phi}}$ &
		& $\mathbf{\omega_{eff}}$ & $q$ &
		\\\hline
		$P_1  $ & $1$ & $0$ &
		$undetermined$ &  & $0$ & $\frac{1}{2}$ \\
		$P_2  $ & $0$ & $1$ &
		$1$& & $1$ & $2$\\
		$P_3  $ & $0$ & $1$ &
		$1$ &  & $1$ & $2$ \\
		$P_4 $ & $0$ & $1$ &
		$-1+\frac{2\beta^{2} m^{2}}{9}$ &  & $-1+\frac{2\beta^{2} m^{2}}{9}$ & $-1+\frac{\beta^{2} m^{2}}{3}$\\
		$P_5 $ & $0$ & $1$ &
		$-1+\frac{2\beta^{2} m^{2}}{9}$ &  & $-1+\frac{2\beta^{2} m^{2}}{9}$ & $-1+\frac{\beta^{2} m^{2}}{3}$\\
		$P_6  $ & $1-\frac{9}{2\beta^{2} m^{2}}$ & $1-\frac{9}{2\beta^{2} m^{2}}$ & $0$ &  & $0$ & $\dfrac{1}{2}$ \\
		$P_7  $ & $1-\frac{9}{2\beta^{2} m^{2}}$ & $1-\frac{9}{2\beta^{2} m^{2}}$ & $0$ &  & $0$ & $\dfrac{1}{2}$ \\
		
		\\\hline\hline
	\end{tabular}
	\label{physical_parameters} \\
	
\end{table}%
%
We now discuss about the local stability of the critical points and to examine the nature of the critical points we have to perturb the system upto first order around the critical points. Also we collect the eigenvalues of the perturb matrix and that describes the stability of the critical points.
\begin{itemize}
\item Critical point $P_1$ exist for all $\beta$ and $m\in [-1,~0) \cup (0,~1]$ in the phase plane $x-y$ and for the point $P_1$ the corresponding physical parameters ${\Omega_{M}}=1$ and ${\Omega_{\phi}}=0$, this imply that the critical point becomes absolutely DM dominated (${\Omega_{M}}=1$) solution for all $\beta$ and $m$. At point $P_1$, eigenvalues of linearized Jacobian matrix are: $$\left\{\lambda_{1(P_1)}=\frac{3}{2},~~~~\lambda_{2(P_1)}= -\frac{3}{2}\right\}$$ 
i.e. they are opposite in sign hence the fixed point is saddle. Parameter $q=\frac{1}{2}$ implies that there always exist a decelerating universe around the critical point $P_1$. So the cosmological scenario of this critical point is it shows decelerated matter dominated (${\Omega_{M}}=1,~{\Omega_{\phi}}=0,~\omega_{eff}=0,~q=\frac{1}{2}$) universe which has saddle like nature.
\item Critical point represents by $P_2$ exist for all $\beta$ and $m\in [-1,~0) \cup (0,~1]$ in the phase plane.
Corresponding physical parameters are given by ${\Omega_{M}}=0,~{\Omega_{\phi}}=1,~\omega_{eff}=1,~\mathbf{\omega_{\phi}}=1$ and $q=2$. The parameters ${\Omega_{M}}=0$ and ${\Omega_{\phi}}=1$ ensures that this critical point is dominated by scalar field DE and also scalar field behaves as stiff fluid as $\omega_{eff}=1$. Another physical parameter $q=2$ suggest that deceleration motion of universe near the critical point is possible. 

Eigenvalues of corresponding linearized Jacobian  matrix for the critical point $P_2$ are: 
$$\left\{\lambda_{1(P_2)}=3,~~~~\lambda_{2(P_2)}=\beta m+3\right\}$$
Therefore the point $P_2$ will be an unstable source if all the eigenvalues are positive and the condition for source is\\
$$[ \beta<-\frac{3}{m},~~-1\leq m<0],~~ \mbox{or}~~ [ \beta>-\frac{3}{m},~~ 0<m\leq1]$$\\
If the eigenvalues are opposite in sign then the fixed point $P_2$ is saddle and in this case the condition becomes\\
$$[ \beta>-\frac{3}{m},~~-1\leq m<0],~~ \mbox{or}~~ [ \beta<-\frac{3}{m},~~ 0<m\leq1]$$\\

\item The point $P_3$ exist for all parameters and corresponding physical parameters are ${\Omega_{M}}=0$, ${\Omega_{\phi}}=1$, $\omega_{eff}=1$, $\mathbf{\omega_{\phi}}=1$ and $q=2$. Corresponding physical parameters ${\Omega_{M}}=0$ and ${\Omega_{\phi}}=1$ ensures the fact this is dominated by scalar field DE and also $\omega_{eff}=1$ and $q=2$ suggest us there always exist a deceleration motion of universe at the critical point $P_3$. Eigenvalues of the perturbed linearized Jacobian matrix for this critical point are: 
$$\left\{\lambda_{1(P_2)}=3,~~~~\lambda_{2(P_2)}=-\beta m+3\right\}$$\\
The point $P_3$ is an unstable source in the phase plane for the following conditions:\\
$$[ \beta>\frac{3}{m},~~-1\leq m<0],~~ \mbox{or}~~ [ \beta<\frac{3}{m},~~ 0<m\leq1]$$\\
And the condition reduces to\\
$$[ \beta<\frac{3}{m},~~-1\leq m<0],~~ \mbox{or}~~ [ \beta>\frac{3}{m},~~ 0<m\leq1]$$\\
for $P_3$ behaves like a saddle. 


\item The nature of the critical points $P_4$, $P_5$ are same in all respect. They exist for the following parameters restrictions: $\left\{-1\leq m<0, ~~ \beta<-\frac{3}{m},~~\beta>\frac{3}{m}\right\}$ ~~\mbox{and}~~ \\ $\left\{0<m\leq1,~~-\frac{3}{m}<\beta<\frac{3}{m}\right\}.$ The points correspond to completely DE dominated solutions in the phase plane where DE behaves as perfect fluid model. Eigenvalues of linearised Jacobian matrix at the critical points are:
$$\left\{\lambda_{1(P_4,P_5)}=-3+\frac{\beta^{2} m^{2}}{3}~~~~~\mbox{and}~~~~~~\lambda_{2(P_4,P_5)}=-3+\frac{2\beta^{2} m^{2}}{3}\right\}$$

The points are hyperbolic in nature since none of the eigenvalues is zero. Linear stability theory is sufficient to find the nature (stability) of the critical points. The points $P_4$, $P_5$ are stable for the following parameters restrictions: $m\in[-1, 0)\cup(0, 1]$ and $[-\frac{3}{\sqrt{2}}<m\beta<0~~\mbox{or}~~ 0<m\beta<\frac{3}{\sqrt{2}}].$ The points behave as saddle like solutions in the phase plane for the following conditions:\\
\begin{enumerate}
	\item $-1\leq m<0,~~
	-3<m\beta<\frac{3}{\sqrt{2}}$
	
	\item $ 0<m\leq 1,~~
	\frac{3}{\sqrt{2}}<m\beta<3$
\end{enumerate}

The points depict the late time evolution of the universe in quintessence region for $0<\beta^2 m^2<3.$ But it never behaves as phantom fluid. Acceleration is possible near the critical points for the region: $\beta^2 m^2<3$. On the other hand, late time evolution is attracted by cosmological constant for $\beta=0$ (i.e., with $\omega_{eff}=q=-1$). Here, DE behaves also as cosmological constant. In this case, the evolution is completely governed by the potential energy of the scalar field. Therefore, from the Friedmann constraint (\ref{Friedmann}), one obtains the exponential expansion of the Universe. Interesting to note that the points describe late time de Sitter accelerated attractor solutions with $\Omega_{\phi}=1$, ${\Omega_{M}}=0$ and in this case DE behaves like cosmological constant fluid as $\omega_{eff}=-1$ for the parameter value $\beta=0$.

\item Finally the critical points $P_6$, $P_7$ exist for $\left\{-1\leq m<0, ~~ \beta\leq-\frac{3}{\sqrt{2}m}~\right\}$ ~~\mbox{and}~~\\ $\left\{0<m\leq1,~\beta\geq\frac{3}{\sqrt{2}m}\right\}$ in the phase plane and they are same in all respect. Corresponding physical parameters are given by $\Omega_{M}=1-\frac{9}{2m^{2}\beta^{2}}$, $\Omega_{\phi}=\frac{9}{2m^{2}\beta^{2}}$, $\omega_{eff}=0$, $\mathbf{\omega_{\phi}}=0$ and $q=\frac{1}{2}$. This shows that the DE connected to the set of points $P_6$ and $P_7$ behaves as any perfect fluid nature depending on $\beta$ and $m$. Eigenvalues of linearized Jacobian matrix near these critical points are:

$$\left\{\lambda_{1(P_6,P_7)}= -\frac{3}{4} +\frac{3}{4} \sqrt{-7+\frac{36}{\beta^{2} m^{2}}},~~~~\lambda_{2(P_6,P_7)}=-\frac{3}{4} -\frac{3}{4} \sqrt{-7+\frac{36}{\beta^{2} m^{2}}}   \right\}$$
and the condition for the critical points to be stable is given by:\\ $\left\{-1\leq m<0, ~~-\frac{6}{\sqrt{7}m}\leq\beta<-\frac{3}{\sqrt{2}m}~\right\}$ ~~\mbox{and}~~ $\left\{0<m\leq1,~\frac{3}{\sqrt{2}m}<\beta\leq\frac{6}{\sqrt{7}m}\right\}$.

\end{itemize}
Cosmological evolution of the critical points extracted from this non-interacting model are presented in the fig.(\ref{noninteracting}) and the evolution of the cosmological parameters are shown in fig.(\ref{noninteracting evolution}) for the parameter values $m=0.998,~\beta=0.001$.
\clearpage 

\subsection{Interacting model}\label{interacting}
In this section we shall study the model with invoking an interaction term $Q$ in the dark sectors. We shall consider the form of $Q$ as \cite{Boehmer2008,Chen2009}
\begin{equation}\label{interaction}
Q=\alpha H \rho_{M}
\end{equation}
where $\alpha$ is a dimensionless coupling of interaction which measures the strength of interaction between the dark sectors. For $\alpha>0$, the energy transfer is occurred from DE to DM while the negative coupling ($\alpha<0$) refers to the energy flow in reverse direction, i.e., from DM to DE.  However, positive coupling is compatible to the second law of thermodynamics and this is required to solve the coincidence problem also. Hence, it is very reassuring to consider the positive coupling in addressing the coincidence problem. On the other hand, negative coupling indicates the decay of DM into DE and such mechanism allows us to consider that the DE is absent in very early universe and it condenses then as a result of slow decay of DM particles \cite{Costa2014}. Although the negative coupling cannot solve the coincidence problem, appearing in observed data fittings, models with it show the most significant departure from zero-coupling \cite{Boehmer2008}. \\ 

The interaction is also well motivated from mathematical point of view. Due to its simplicity, the system of ordinary differential equations remains two dimensional because cosmological variable $H$ can be eliminated. Now, by using the term (\ref{interaction}) in the system of ODEs (\ref{non-autonomous}), we obtain the 2D autonomous system in $x,~y$:
\begin{eqnarray}\label{autonomous Interacting}
\begin{split}
	\frac{dx}{dN}& =\frac{3}{2} x \left(x^2-y^2-1\right)-\beta m y^2-\frac{\alpha}{2x} (1-x^2 -y^2),& \\
	\frac{dy}{dN}& =y \left(\beta m x+\frac{3}{2} \left(x^2-y^2+1\right)\right),
	&~~
\end{split}
\end{eqnarray}
where $\beta$ is a constant and it can take any real value for particular choices of $V(\phi)$ and $\lambda(\phi)$ which we already have discussed in the beginning of this section. We also consider $m= \frac{  \sqrt{3} \sqrt{2 n+1} }{n+2}$ and thus $m$ is bounded as $-1\leq m,~~ m\leq1, ~~m\neq 0$. The autonomous system (\ref{autonomous Interacting}) has a singularity at $x=0$ and to remove this, one has to consider the system with all equations multiplied by the positive definite term $x^2$. The dynamics of new  system will remain same as before. In fact, one can get the critical points at limiting case. This method has already been used earlier in the literature. Therefore, after multiplying by $x^2$ to the terms in right hand sides of both the equations, the system (\ref{autonomous Interacting}) reduces to the form:
\begin{eqnarray}\label{autonomous Interacting singularity free}
\begin{split}
	\frac{dx}{dN}& =\frac{3}{2} x^3 \left(x^2-y^2-1\right)-\beta m x^2 y^2-\frac{\alpha}{2} x (1-x^2 -y^2),& \\
	\frac{dy}{dN}& =x^2 y  \left(\beta m x+\frac{3}{2} \left(x^2-y^2+1\right)\right),
	&~~
\end{split}
\end{eqnarray}

\subsubsection{Phase plane analysis}
We shall now analyse the critical points of the autonomous system (\ref{autonomous Interacting}) by equating the right part of both the equations to zero. The system admits nine critical points in $x-y$ phase plane are as follows
\begin{itemize}
\item Set of critical Points $C_1:~(0,~~y_c)$
\item Critical Points $C_{2}, C_{3}:~(\pm 1,~~0)$

\item Citical Points $C_4,~ C_5:~(\pm \sqrt{-\frac{\alpha}{3}},~~0)$

\item Critical Points $C_6,~C_7:~\left(-\frac{\beta m }{3},~\pm  \sqrt{1-\frac{\beta^{2}m^{2}}{9}}\right)$

\item Critical Point $C_8,~C_9:~\left( \frac{\alpha-3}{2\beta m} ,~  \pm \frac{\sqrt{\frac{4}{3}\alpha \beta^{2} m^{2}+(\alpha-3)^{2}}}{2\beta m} \right)$
\end{itemize}
The points $C_4$ and $C_5$ exist only for negative coupling ($\alpha<0$) in the phase plane. The critical points and the corresponding cosmological parameters are displayed in the table \ref{physical_parameters int} and the eigenvalues of linearised Jacobian matrix are presented in the table \ref{eigenvalues int} 

\begin{table}[h!]  \centering \fontsize{9.6pt}{5pt}
\caption{The critical points and the corresponding cosmological parameters.}%
\begin{tabular}
	[c]{cccccccc}\hline\hline
	\textbf{Critical Point}&$\mathbf{\Omega_{M}}$& $\mathbf{\Omega_{\phi}}$ & $\mathbf{\omega_{\phi}}$ &
	& $\mathbf{\omega_{eff}}$ & $q$ &
	\\ \hline \\
	$C_1  $ & $1-y_c^2$ & $y_c^2$ &
	$-1$ &  & $-y_c^2$ & $\frac{1}{2}-\frac{3y_c^2}{2}$ \\ \\
	$C_2,~C3  $ & $0$ & $1$ &
	$1$& & $1$ & $2$\\ \\
	$C_4,~C_5  $ & $1+\frac{\alpha}{3}$ & $-\frac{\alpha}{3}$ &
	$1$ &  & $-\frac{\alpha}{3}$ & $\frac{1}{2}-\frac{\alpha}{2}$ \\ \\
	$C_6,~C_7 $ & $0$ & $1$ &
	$-1+\frac{2\beta^{2} m^{2}}{9}$ &  & $-1+\frac{2\beta^{2} m^{2}}{9}$ & $-1+\frac{\beta^{2} m^{2}}{3}$\\ \\
	$C_8,~C_9  $ & $\frac{\left(3-\alpha \right) \left(\beta^2 m^2 -\frac{3}{2}(3-\alpha) \right)}{3\beta^{2} m^{2}}$ & $\frac{2\alpha \beta^2 m^2 +3(\alpha-3)^2}{6\beta^{2} m^{2}}$ & $-\frac{2\alpha \beta^2 m^2}{2\alpha \beta^2 m^2 +3(\alpha-3)^2}$ &  & $-\frac{\alpha}{3}$ & $\dfrac{1}{2}(1-\alpha)$ \\
	
	\\\hline\hline
\end{tabular}
\label{physical_parameters int} \\

\end{table}%
%

\begin{table}[h!]  \centering \fontsize{8.5pt}{5pt}
\caption{The eigenvalues for critical points where $\varXi=16\alpha^2 \beta^6 m^6 -24m^4 \alpha(\alpha-3)(\alpha-15)\beta^4 -135m^2 (\alpha-3)^2 (\alpha-7)(\alpha-\frac{3}{5}) \beta^2 -108(\alpha-3)^5$}%
\setlength{\tabcolsep}{0.05cm}
\renewcommand{\arraystretch}{}
\begin{tabular}
[c]{cccc}\hline\hline
\textbf{Critical Point}&$\mathbf{\lambda_1}$& $\mathbf{\lambda_2}$ & 
\\\hline
$C_1  $ & $0$ & $\frac{1}{2}\alpha y_c^2 -\frac{1}{2}\alpha$ &
\\ \\
$C_2  $ & $\beta m+3$ & $3+\alpha$ &
\\ \\
$C_3  $ & $-\beta m+3$ & $3+\alpha$ &
\\ \\
$C_4 $ & $\frac{1}{3}\alpha^2 +\alpha$ & $\frac{\sqrt{3}}{9} (-\alpha)^{\frac{3}{2}} \beta m+\frac{\alpha^2}{6}-\frac{\alpha}{2}$ &
\\ \\
$C_5 $ & $\frac{1}{3}\alpha^2 +\alpha$ & $-\frac{\sqrt{3}}{9} (-\alpha)^{\frac{3}{2}} \beta m+\frac{\alpha^2}{6}-\frac{\alpha}{2}$ &
\\ \\
$C_6,~C_7  $ & $\frac{\beta^2 m^2}{27} (\beta^2 m^2-9)$ & $\frac{\beta^2 m^2}{27} (2\beta^2 m^2+3\alpha-9)$ &  \\ \\
$C_8,~C_9  $ & $-\frac{1}{12m^3 \beta^3}\left[(\alpha-3) \left ( -\frac{\sqrt{\varXi}}{4} +m \beta \left(\alpha \beta^2 m^2 +\frac{9(\alpha+1)(\alpha-3)}{4} \right) \right )  \right]$ & $-\frac{1}{12m^3 \beta^3}\left[(\alpha-3) \left ( \frac{\sqrt{\varXi}}{4} +m \beta \left(\alpha \beta^2 m^2 +\frac{9(\alpha+1)(\alpha-3)}{4} \right) \right )  \right]$ &  \\
\\\hline\hline
\end{tabular}
\label{eigenvalues int} \\

\end{table}%
%

\begin{itemize}

\item Set of critical points $C_1$ exists only for $0\leq y_c^2 \leq1$ in the phase plane. Therefore, its existence is independent of $\alpha,~\beta$ and the anisotropic parameter $m$. The set is the solution with combination of both the DE and DM where DE behaves as cosmological constant-like fluid since $\omega_{\phi}=-1$. The set of critical points having exactly one vanishing eigenvalue for $\alpha\neq 0,~~y_c\neq 1$ (see in table \ref{eigenvalues int}), is the normally hyperbolic set. To check the stability of this set, it is sufficient to examine the sign of the remaining non-zero eigenvalue. The set will be stable in the phase plane for $\alpha>0,~~-1<y_c<1$.
The stability of this set is shown numerically in the fig. (\ref{scalingC1}) for $m=0.9986,~\alpha=0.5,~\beta=0.5$.
The expansion of the Universe will be accelerating in the interval $y_c^2>\frac{1}{3}$.
The set becomes completely DE dominated for $|y_c|\longrightarrow 1$. On the other hand, it will be DM dominated solution for $y_c\longrightarrow 0$. Furthermore, for $y_c=0$, a decelerated expansion is exhibited ($ q=\frac{1}{2}$) by a specific point ($0,~0$) on the set $C_1$. Here, a dust dominated ($\omega_{eff}=0$) decelerated phase of the evolution is obtained. However, for $y_c=0$, the eigenvalues associated with this point are $\lambda_1=0,~\lambda_2=-\frac{1}{2}\alpha$. As a result, the point will be non-hyperbolic in nature. However, for $\alpha>0$, the set will become a solution in late time cosmology since there exists a one dimensional stable manifold in one eigen direction. In this case, the point (0,0) may predict the future decelerated dust dominated universe. 
\begin{figure}
\centering
\subfigure[]{%
\includegraphics[width=10cm,height=10cm]{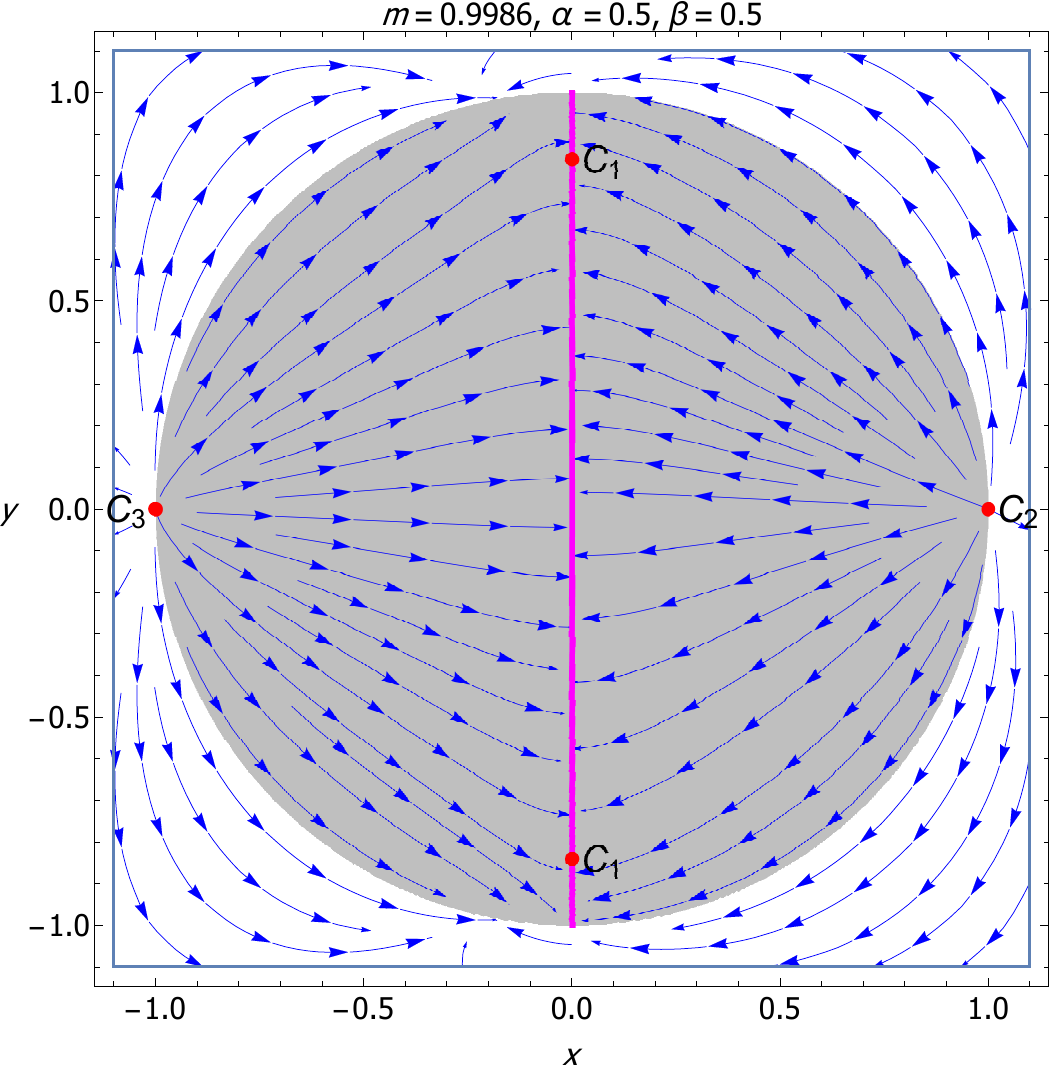}\label{scalingC1}}
\qquad
\subfigure[]{%
\includegraphics[width=11cm,height=8cm]{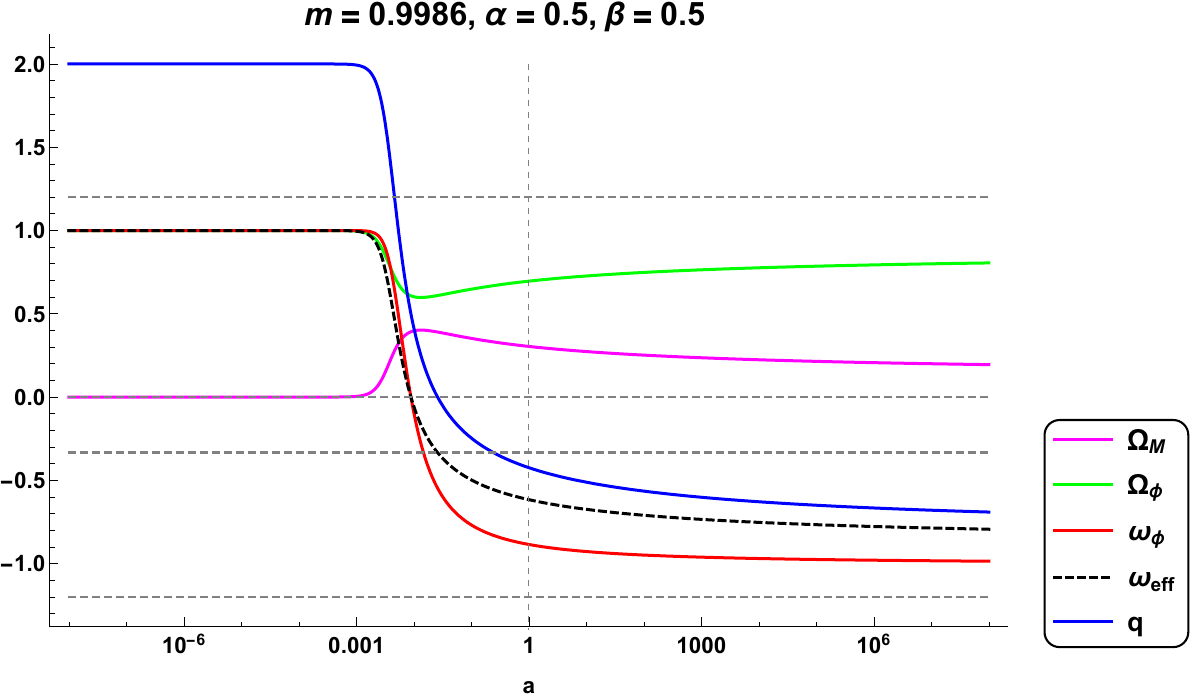}\label{evolutionC1}}

\caption{The figure shows the phase plane on $x-y$ plane and the evolution of cosmological parameters for $m=0.9986$, $\alpha=0.5$ and $\beta=0.5$. In panel (a) late time scaling attractor $C_1$ is shown and is supported by present observations  and panel (b) shows that late time accelerated evolution of the universe is attracted in quintessence region .}
\label{coupled C1 scaling}
\end{figure}
\clearpage

\item The points $C_2, ~ C_3$ exist for all parameters. These points are completely DE dominated solutions ($\Omega_{\phi}=1$) and the DE behaves as stiff fluid here ($\omega_{\phi}=1$). These points always exhibit decelerated expansion of the Universe ($q=2$). Eigenvalues for the critical point $C_2$ are $\lambda_1=\beta m+3,~~\lambda_2=3+\alpha$. The stability of this points depend on the parameters involved in eigenvalues. The condition for stability of the point $C_2$ is $$[\alpha<-3,~~\beta>-\frac{3}{m},~~-1\leq m<0],~~ \mbox{or}~~ [\alpha<-3,~~\beta<-\frac{3}{m},~~ 0<m\leq1]$$
The point $C_2$ will be an unstable source if all the eigenvalues are positive and the condition for source is
$$[\alpha>-3,~~\beta<-\frac{3}{m},~~-1\leq m<0],~~ \mbox{or}~~ [\alpha>-3,~~\beta>-\frac{3}{m},~~ 0<m\leq1].$$
The eigenvalues of the critical point $C_3$ are $\lambda_1=-\beta m+3,~~\lambda_2=3+\alpha$.	
However, the critical point $C_3$ has stability in the following parameter restrictions:
$$[\alpha<-3,~~\beta<\frac{3}{m},~~-1\leq m<0],~~ \mbox{or}~~ [\alpha<-3,~~\beta>\frac{3}{m},~~ 0<m\leq1]$$	
We also obtain the condition for the critical point $C_3$ to be unstable source in the phase plane is that
$$[\alpha>-3,~~\beta>\frac{3}{m},~~-1\leq m<0],~~ \mbox{or}~~ [\alpha>-3,~~\beta<\frac{3}{m},~~ 0<m\leq1].$$	
\item The points $C_4,~ C_5$ exist for $\alpha \in (-3,0)$. These points represent the combination of both the DE and DM where the DE behaves as stiff fluid in nature. Acceleration is not possible for these points. Eigenvalues of the linearised Jacobian matrix at the critical point $C_4$ is: 
$\lambda_1= \frac{1}{3}\alpha^2 +\alpha$,~~  $\lambda_2= \frac{\sqrt{3}}{9} (-\alpha)^{\frac{3}{2}} \beta m+\frac{\alpha^2}{6}-\frac{\alpha}{2}.$ 
This point is stable for the following conditions:\\
$\left[-3<\alpha<0, ~~\mbox{and}~~\left\{2\beta> \frac{(\alpha-3)}{m} \sqrt{-\frac{3}{\alpha}},~~-1\leq m <0 \right\} \mbox{or}~~\{ 2\beta<- \frac{(\alpha-3)}{m} \sqrt{-\frac{3}{\alpha}},~~0< m \leq1    \} \right]$.  	  
The eigenvalues for the critical point $C_5$ are
$\lambda_1= \frac{1}{3}\alpha^2 +\alpha$,~~  $\lambda_2= -\frac{\sqrt{3}}{9} (-\alpha)^{\frac{3}{2}} \beta m+\frac{\alpha^2}{6}-\frac{\alpha}{2}.$
This point can be stable in the phase plane for:\\
$\left[-3<\alpha<0, ~~\mbox{and}~~\left\{2\beta<- \frac{(\alpha-3)}{m} \sqrt{-\frac{3}{\alpha}},~~-1\leq m <0 \right\} \mbox{or}~~\{ 2\beta> \frac{(\alpha-3)}{m} \sqrt{-\frac{3}{\alpha}},~~0< m \leq1    \} \right]$.
Therefore, the critical points predict the future deceleration of the evolution of the universe for negative coupling of interaction term along with some restricted values of other parameters.
\item Critical points $C_6, ~ C_7$ are same in all respect.\\
They exist for 
$\left\{-1\leq m<0, ~~ \beta<-\frac{3}{m},~~\beta>\frac{3}{m}\right\}$ ~~\mbox{and}~~ $\left\{0<m\leq1,~~-\frac{3}{m}<\beta<\frac{3}{m}\right\}.$
These are completely DE dominated solutions in the phase plane and DE represented by perfect fluid equation of state $\omega_{\phi}=-1+\frac{2\beta^2 m^2}{9}$ (see table \ref{physical_parameters int}). DE can be cosmological constant for $\beta=0$, quintessence for $0<\beta^2 m^2<3$ and any other exotic type fluid for $3<\beta^2 m^2<\frac{9}{2}$ but it can never behave as phantom like fluid. Accelerating universe is achieved near the critical points for the region $\beta^2 m^2<3$.
The points $C_6,~ C_7$ are stable for the following parameters restrictions:	
\begin{enumerate}
\item $-1\leq m<0 :~~\\
(i) ~  \alpha \leq -3~~\mbox{and}~~ \left(\frac{3}{m}<\beta <0 ~~\mbox{or}~~ 0<\beta <-\frac{3}{m}\right), ~~\mbox{or}~~\\
(ii) -3<\alpha <3 ~~\mbox{and}~~ \left(-\sqrt{\frac{3}{2}} \sqrt{-\frac{\alpha -3}{m^2}}<\beta <0 ~~\mbox{or}~~ 0<\beta <\sqrt{\frac{3}{2}} \sqrt{-\frac{\alpha -3}{m^2}}\right)$

\item $ 0<m\leq 1 :~~\\
(i) ~  \alpha \leq -3~~\mbox{and}~~ \left(-\frac{3}{m}<\beta <0 ~~\mbox{or}~~ 0<\beta <\frac{3}{m}\right), ~~\mbox{or}~~\\
(ii) -3<\alpha <3 ~~\mbox{and}~~ \left(-\sqrt{\frac{3}{2}} \sqrt{-\frac{\alpha -3}{m^2}}<\beta <0 ~~\mbox{or}~~ 0<\beta <\sqrt{\frac{3}{2}} \sqrt{-\frac{\alpha -3}{m^2}}\right)$
\end{enumerate}
Therefore, the critical points depict the late time accelerated which represent the future attractors in the quintessence region:
$-1\leq m<0,~0<m\leq1$	and \\
 $\bigg[ \left(\alpha\leq1,~~0<\beta<\sqrt{3} \sqrt{\frac{1}{m^2}},~~ \mbox{or}~~ -\sqrt{3} \sqrt{\frac{1}{m^2}} <\beta<0  \right)          	
~\mbox{or}~ \\
\left( 1<\alpha<3,~~0<\beta<\frac{\sqrt{3}}{\sqrt{2}} \sqrt{\frac{3-\alpha}{m^2}},~~ \mbox{or},~~
-\sqrt{\frac{3}{2}} \sqrt{\frac{3-\alpha}{m^2}}<\beta<0 \right)\bigg] $
\begin{figure}
\centering
\subfigure[]{%
\includegraphics[width=10cm,height=10cm]{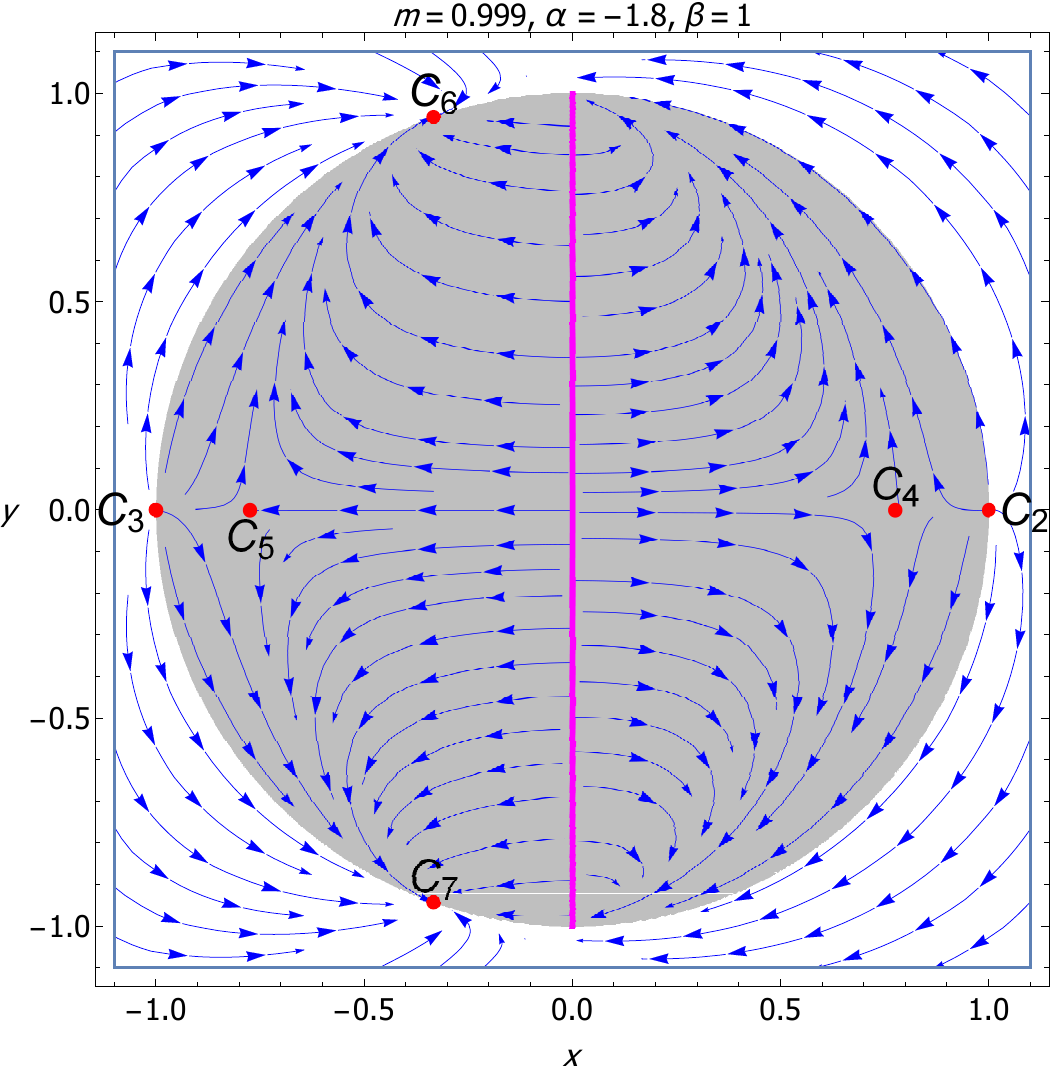}\label{C6C7stable}}

\caption{The figure shows the phase plane of the autonomous system (\ref{autonomous Interacting singularity free}) on the x-y plane where the parameter values are $m=0.999$, $\alpha=-1.8$ and $\beta=1$. Here, critical points $C_6$ and $C_7$ are scalar field dominated late time attractors, the points $C_4$ and $C_5$ are saddle like solutions and the points $C_2$ and $C_3$ are unstable nodes in the phase plane. }
\label{coupled C6C7}
\end{figure}
\clearpage 

Critical points $C_6$ and $C_7$ have another interesting feature when the DE of them behave like dust. In particular, if we consider $\beta^2 m^2 =\frac{9}{2}$, the DE equation of state becomes $\omega_{\phi}=0$ and consequently $\omega_{eff}=0$ and $q=\frac{1}{2}$ which mimics the scenario as if the universe is dust dominated. Depending on restriction of $\alpha$, the points can be saddle like solution. Therefore, the points may be mimicking as transient evolution of the universe. 

\item The points $C_8,~ C_9$ are same in nature in all respect. They exist for the following conditions:

\begin{enumerate}
\item 	$-1\leq m<0: ~~\\
(i)~ \alpha =-3 ~~\mbox{and}~~ \left(\beta =\frac{3}{m}~~\mbox{or}~~ \beta= -\frac{3}{m} \right), ~~\mbox{or}~~\\
(ii)~ -3<\alpha <0 ~~\mbox{and}~~ \left(-\frac{\sqrt{3} (\alpha -3)}{2 \sqrt{-\alpha } m}\leq \beta \leq \frac{\sqrt{\frac{3}{2}} \sqrt{3-\alpha }}{m}~~\mbox{or}~~ -\frac{\sqrt{\frac{3}{2}} \sqrt{3-\alpha }}{m}\leq \beta \leq \frac{\sqrt{3} (\alpha -3)}{2 \sqrt{-\alpha } m}\right), ~~\mbox{or}~~\\
(iii)~ 0\leq \alpha <3 ~~\mbox{and}~~ \left(2 \beta \leq \frac{\sqrt{6} \sqrt{3-\alpha }}{m}~~\mbox{or~~} 2 \beta +\frac{\sqrt{6} \sqrt{3-\alpha }}{m}\geq 0\right), ~~\mbox{or}~~\\
(iv)~ \alpha =3 ~~\mbox{and}~~ \beta \neq 0$

\item  $0<m\leq 1: ~~\\
(i)~ \alpha =-3 ~~\mbox{and}~~ \left(\beta =\frac{3}{m}~~\mbox{or}~~ \beta= -\frac{3}{m} \right), ~~\mbox{or}~~\\
(ii)~ -3<\alpha <0 ~~\mbox{and}~~ \left(\frac{\sqrt{3} (\alpha -3)}{2 \sqrt{-\alpha } m}\leq \beta \leq- \frac{\sqrt{\frac{3}{2}} \sqrt{3-\alpha }}{m}~~\mbox{or}~~ \frac{\sqrt{\frac{3}{2}} \sqrt{3-\alpha }}{m}\leq \beta \leq -\frac{\sqrt{3} (\alpha -3)}{2 \sqrt{-\alpha } m}\right), ~~\mbox{or}~~\\
(iii)~ 0\leq \alpha <3 ~~\mbox{and}~~ \left(2 \beta \geq \frac{\sqrt{6} \sqrt{3-\alpha }}{m}~~\mbox{or~~} 2 \beta +\frac{\sqrt{6} \sqrt{3-\alpha }}{m}\leq 0\right), ~~\mbox{or}~~\\
(iv)~ \alpha =3 ~~\mbox{and}~~ \beta \neq 0$
\end{enumerate}
These points correspond to solutions with combination of both DE and DM. The ratio of DE and DM is $r=\frac{\Omega_{\phi}}{\Omega_{M}}=\frac{2\alpha \beta^2 m^2 +3(\alpha-3)^2}{2(3-\alpha)(\beta^2 m^2-\frac{3}{2}(3-\alpha))}$

These points will become completely DE dominated for $\alpha=3$ ($\Omega_{M}=0,~\Omega_{\phi}=1$ see table \ref{physical_parameters int}). The DE behaves as any perfect fluid nature ($\omega_{\phi}=-\frac{2\alpha \beta^2 m^2}{2\alpha \beta^2 m^2 +3(\alpha-3)^2}$) which can be quintessence for $\beta^2 m^2 >\frac{3}{4\alpha}(\alpha-3)^2$, cosmological constant for $\alpha=3$ and any other exotic type fluid for $-\frac{3}{2\alpha}(\alpha-3)^2 <\beta^2 m^2 <\frac{3}{4\alpha (\alpha-3)^2}$ but it can never be a phantom fluid. 

Acceleration is possible near the points when $\alpha>1$.

The points $C_8,~ C_9$ are stable for the following conditions:
\begin{enumerate}
\item $-1\leq m<0 : ~~\\ 
(i)~ -3<\alpha <0 ~~\mbox{and}~~ \left(-\frac{\sqrt{3} (\alpha -3)}{2 \sqrt{-\alpha } m}<\beta <\frac{\sqrt{\frac{3}{2}} \sqrt{3-\alpha }}{m} ~~\mbox{or}~~ -\frac{\sqrt{\frac{3}{2}} \sqrt{3-\alpha }}{m}<\beta <\frac{\sqrt{3} (\alpha -3)}{2 \sqrt{-\alpha } m}\right), ~~\mbox{or}~~\\
(ii)~ \alpha =0 ~~\mbox{and}~~ \left(\beta <\frac{3}{\sqrt{2} m} ~~\mbox{or}~~ \beta >-\frac{3}{\sqrt{2} m}\right),~~\mbox{or}~~\\
(iii)~ 0<\alpha <3 ~~\mbox{and}~~ \left(\frac{3 \sqrt{-\alpha +\frac{3}{\alpha }+2}}{2 m}<\beta <\frac{\sqrt{\frac{3}{2}} \sqrt{3-\alpha }}{m} ~~\mbox{or}~~ -\frac{\sqrt{\frac{3}{2}} \sqrt{3-\alpha }}{m}<\beta <-\frac{3 \sqrt{-\alpha +\frac{3}{\alpha }+2}}{2 m}\right)$

\item  $0<m\leq 1 : ~~\\ 
(i)~ -3<\alpha <0 ~~\mbox{and}~~ \left(\frac{\sqrt{3} (\alpha -3)}{2 \sqrt{-\alpha } m}<\beta <-\frac{\sqrt{\frac{3}{2}} \sqrt{3-\alpha }}{m} ~~\mbox{or}~~ \frac{\sqrt{\frac{3}{2}} \sqrt{3-\alpha }}{m}<\beta <-\frac{\sqrt{3} (\alpha -3)}{2 \sqrt{-\alpha } m}\right), ~~\mbox{or}~~\\
(ii)~ \alpha =0 ~~\mbox{and}~~ \left(\beta <-\frac{3}{\sqrt{2} m} ~~\mbox{or}~~ \beta >\frac{3}{\sqrt{2} m}\right),~~\mbox{or}~~\\
(iii)~ 0<\alpha <3 ~~\mbox{and}~~ \left(-\frac{3 \sqrt{-\alpha +\frac{3}{\alpha }+2}}{2 m}<\beta <-\frac{\sqrt{\frac{3}{2}} \sqrt{3-\alpha }}{m} ~~\mbox{or}~~ \frac{\sqrt{\frac{3}{2}} \sqrt{3-\alpha }}{m}<\beta <\frac{3 \sqrt{-\alpha +\frac{3}{\alpha }+2}}{2 m}\right)$
\end{enumerate}
The points represent the late time accelerated evolution of the universe attracted in quintessence era for  

$-1\leq m<0,~~0< m \leq1$ and\\
$ \left\{1<\alpha<3,~~\left( -\frac{3}{2} \sqrt{\frac{3+2\alpha-\alpha^2}{m^2 \alpha}}<\beta<-\sqrt{\frac{3}{2}} \sqrt{\frac{3-\alpha}{m^2}}  \right)~~  \mbox{or}~~ \left( \sqrt{\frac{3}{2}} \sqrt{\frac{3-\alpha}{m^2}} <\beta< \frac{3}{2} \sqrt{\frac{3+2\alpha-\alpha^2}{m^2 \alpha}} \right)      \right \} $

where DE also behaves as quintessence-like fluid. In that case, the late time solutions are scaling and supported by observations and these solutions can give  solutions to alleviate the coincidence problem. It is interesting to note that the points are completely dominated by DE for $\alpha=3$, then these show the potential energy of scalar field dominated de Sitter universe which is accelerating. However, these are not future attractors. In this case the critical points become $(0, ~\pm 1) $\\
These points are unstable source if all the corresponding eigenvalues are positive and the condition for source is
$$[\alpha<3,~~-4\alpha\beta^{2}m^{2}+9(\alpha+1)(\alpha-3)<\frac{\sqrt{\varXi}}{m\beta}<4\alpha\beta^{2}m^{2}+9(\alpha+1)(\alpha-3),~~ m\in [-1,~0) \cup (0,~1]~]$$

\end{itemize}


\begin{figure}
\centering
\subfigure[]{%
\includegraphics[width=7.2cm,height=7.2cm]{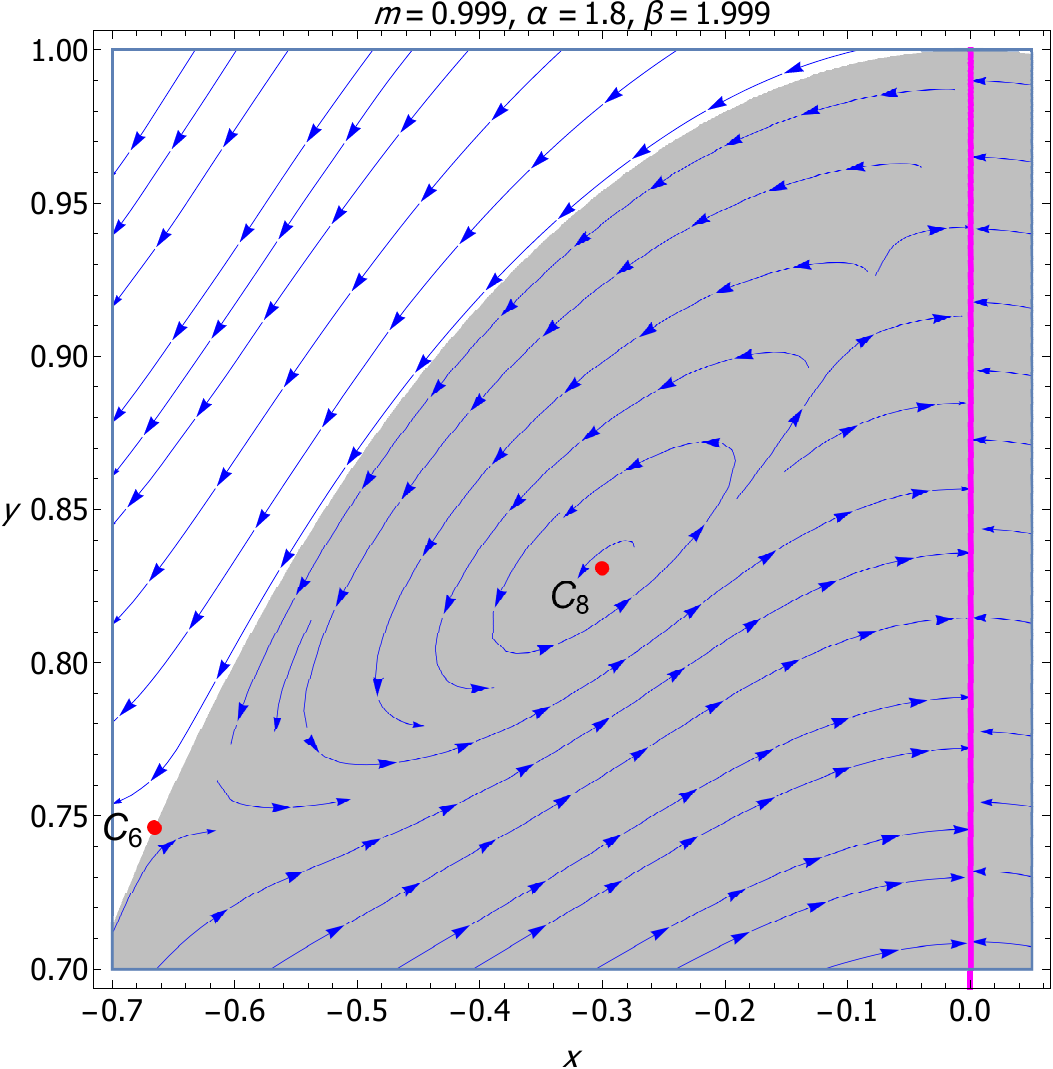}\label{scalingC8}}
\qquad
\subfigure[]{%
\includegraphics[width=7.2cm,height=7.2cm]{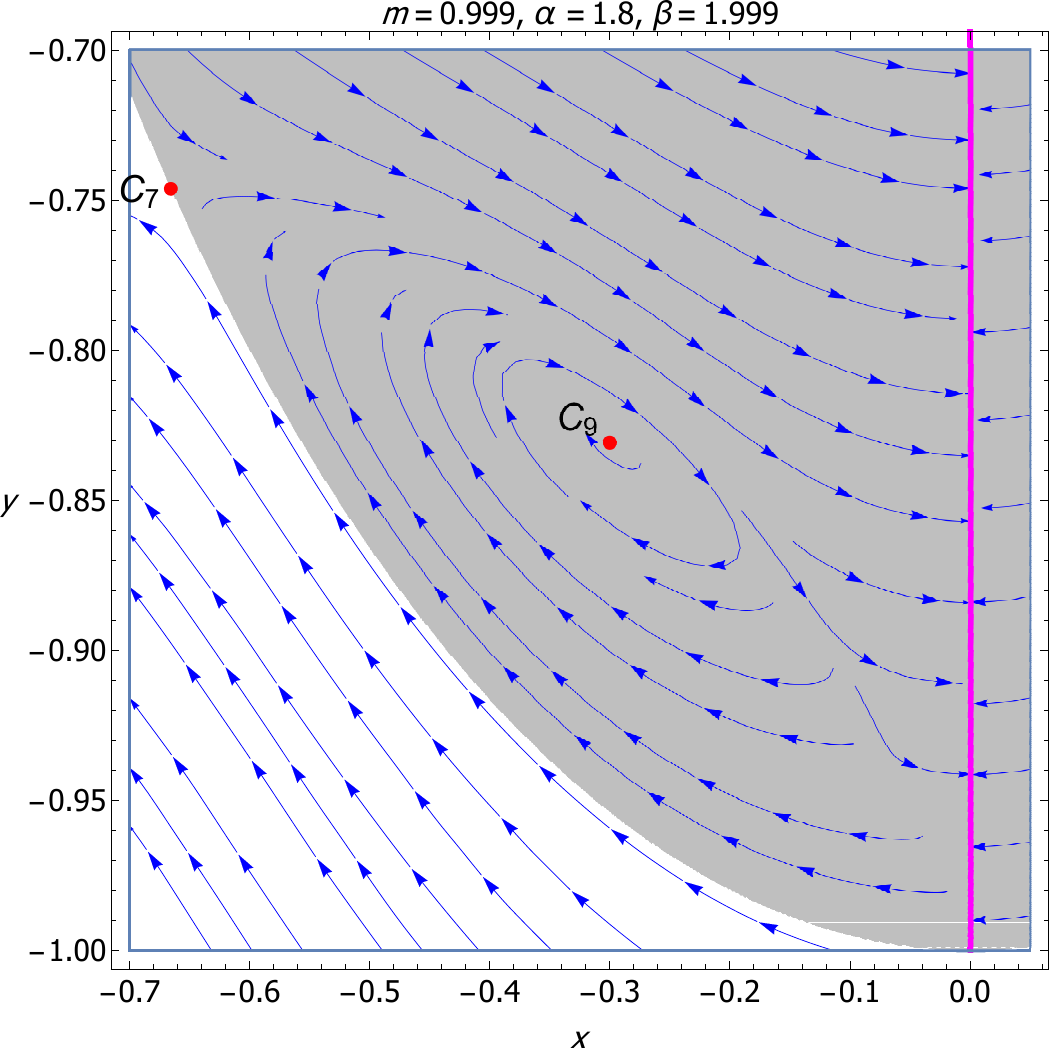}\label{scalingC9}}
\qquad
\subfigure[]{%
\includegraphics[width=12cm,height=7.5cm]{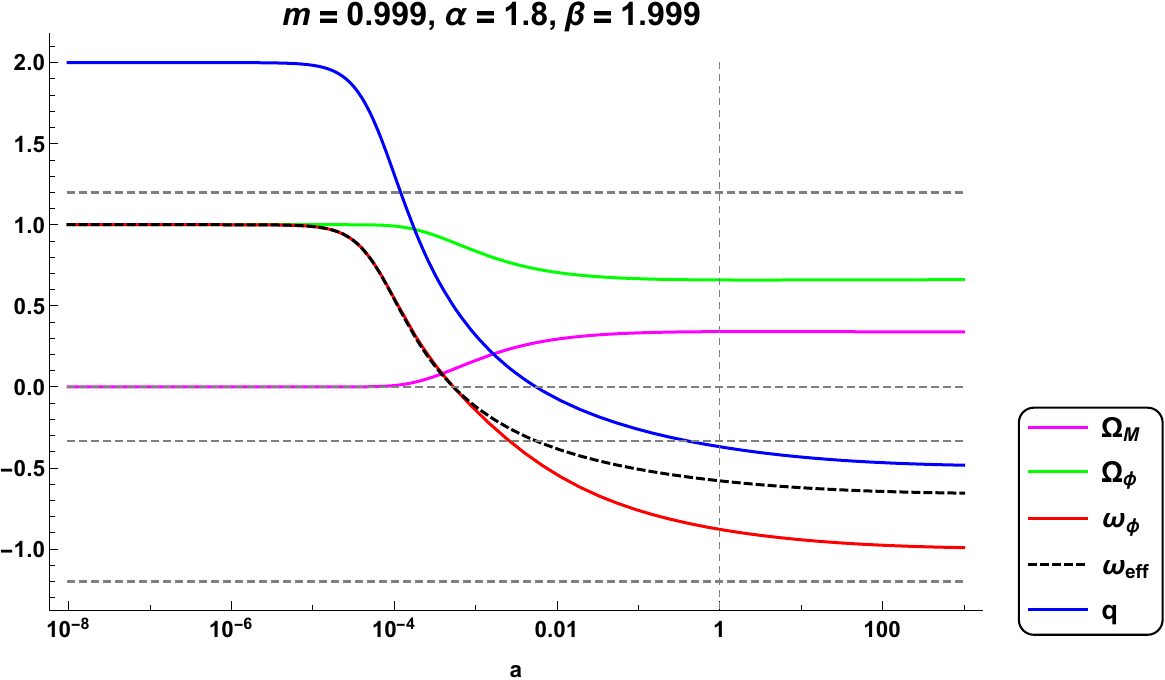}\label{evolutionC8C9}}
\caption{The figure shows the phase plane of the autonomous system (\ref{autonomous Interacting singularity free}) on the x-y plane for the parameter values of $m=0.999$, $\alpha=1.8$ and  $\beta=1.999$. In panel (a) the scaling solution $C_8$ is late time attractor. Panel (b) shows that the solution $C_9$ is scaling attractor in the phase plane. Panel (c) shows the evolution of cosmological parameters. The scalar field dominated universe is evolving in quintessence era at late times }
\label{coupled C8 C9 scaling}
\end{figure}
\clearpage 

\begin{figure}
\centering
\subfigure[]{%
\includegraphics[width=7.2cm,height=7.2cm]{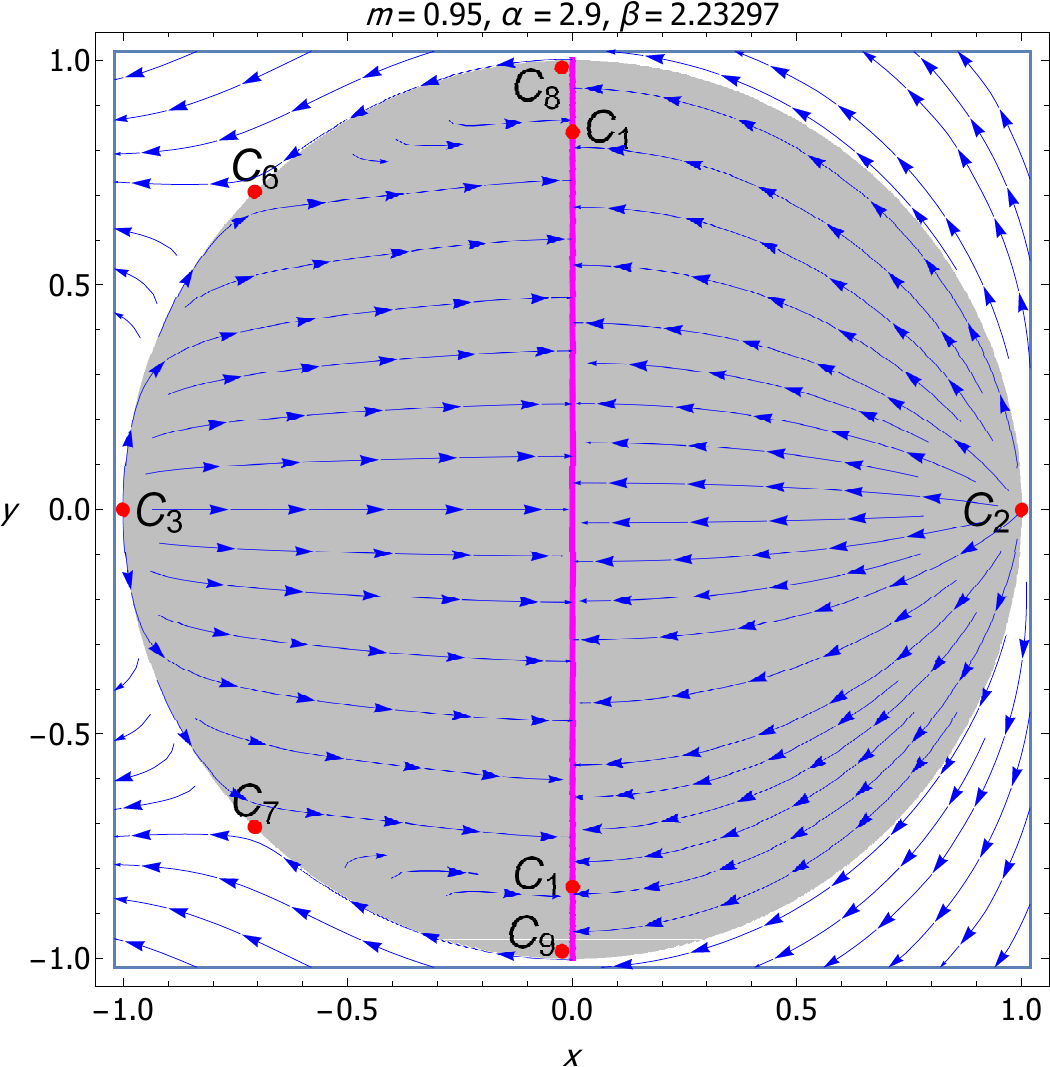}\label{complete scenario1}}
\qquad
\subfigure[]{%
\includegraphics[width=7.2cm,height=7.2cm]{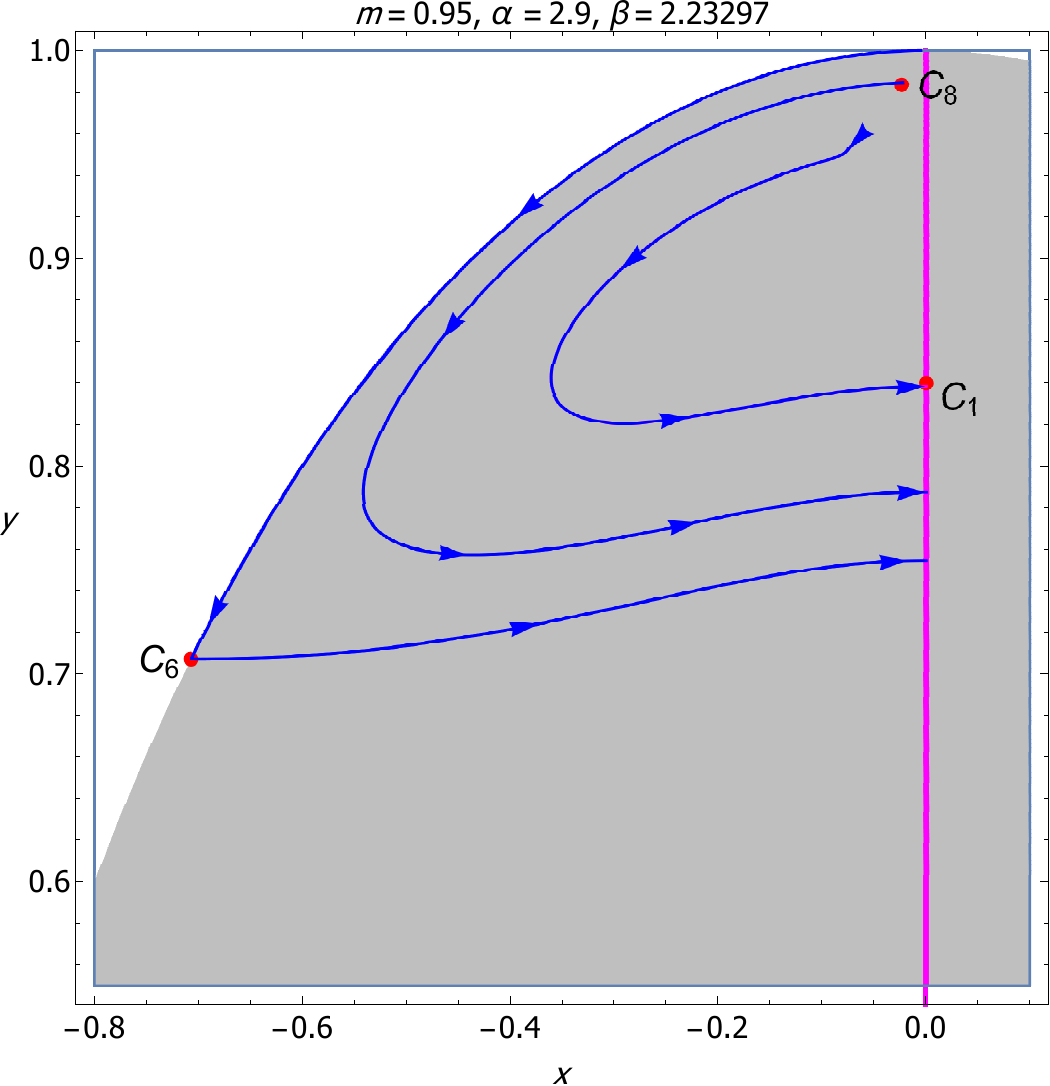}\label{complete scenario2}}
\qquad
\subfigure[]{%
\includegraphics[width=7.2cm,height=7.2cm]{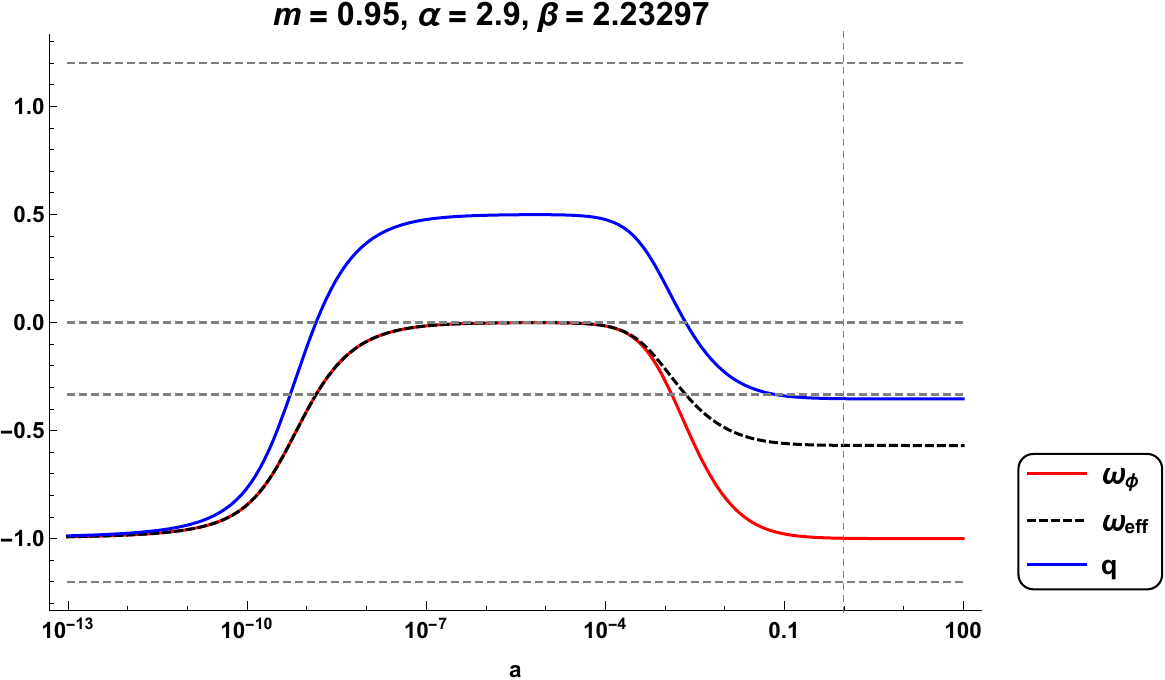}\label{complete scenario evolution}}
\qquad
\subfigure[]{%
\includegraphics[width=7.2cm,height=7.2cm]{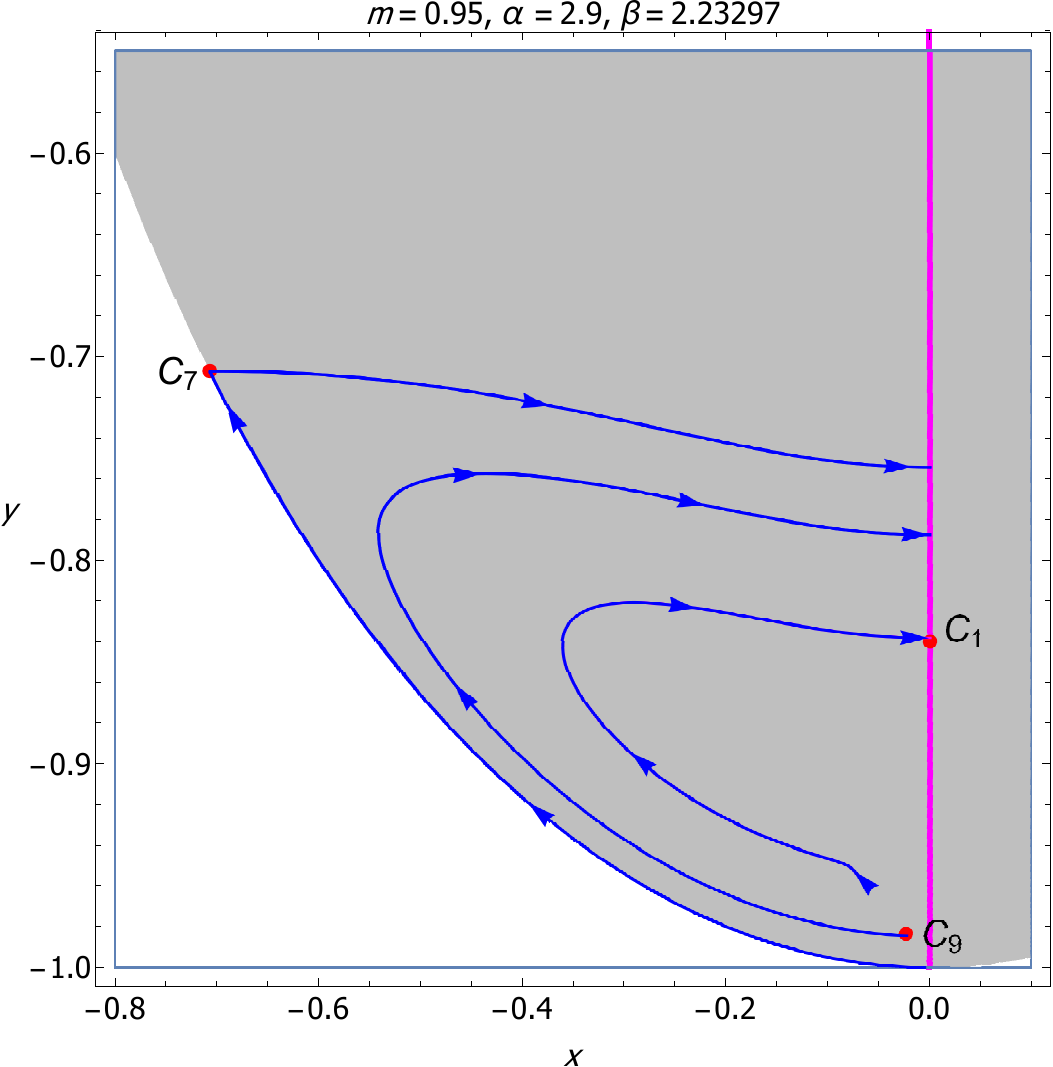}\label{complete scenario3}}
\caption{The figure shows that the phase plane of the autonomous system (\ref{autonomous Interacting singularity free}) on the x-y plane for the parameter values: $m=0.95,~~\alpha=2.9,~~\beta=2.23297$.
In panel (a) the points $C_2$ and $C_3$ are the kinetic energy of scalar field dominated decelerated solutions which are source (unstable nodes in phase plane), points $C_6$ and $C_7$ are dust dominated decelerated solutions representing the intermediate phase where DE behaves as dust.
The points $C_8$ and $C_9$ are scalar field dominated past attractors. The point $C_1$ is stable attractor. In panel (b) a complete evolution of the universe is shown where the point $C_8$ is a source, the point $C_6$ saddle and $C_1$ is a stable solution. The panel
(c) shows the evolution of decelerating parameter $q$, $\omega_{\phi}$ and $\omega_{eff}$  which shows the unified description of universe from early inflation to late time acceleration connecting through intermediate decelerated phase. Panel (d) shows the critical point $C_9$ source, $C_7$ saddle and $C_1$ is a stable attractor.}
\label{coupled complete evolution}
\end{figure}
\clearpage

\section{Physical implications of critical points}\label{sec-cp}

We studied the phase plane analysis of coupled and uncoupled model in the background of spatially flat, LRS anisotropic Bianchi I universe where scalar field taken as DE and pressureless dust as the DM. We first analyzed the uncoupled model of DE, namely the autonomous system (\ref{autonomous noninteracting}). We extracted all possible critical points from this system. We obtained several critical points describing early DM dominated solutions and late time DE dominated accelerated solutions. Also, we measured the anisotropic effect (by the parameter $m$) in the evolution at late-times as well as in past.
Most of the critical points are affected by anisotropic parameter. First of all, a dust dominated decelerated Universe is achieved by the critical point $P_1$. This point is saddle-like solution, so it represents the intermediate phase of the Universe.

The points $P_2$ and $P_3$ are kinetic term of the scalar field dominated decelerated solutions. The stability of the points affected directly by anisotropic term $m$. So, depending upon the parameters, the points may represent the early phase of the universe because the points can be past attractors in some parameters regions. But, unfortunately, the expansion of the Universe is always decelerated.

The points $P_4$ and $P_5$ correspond to late time accelerated evolution of the universe attracted in quintessence regime where the DE behaves as quintessence also. In presence of anisotropic effect the features is also achieved. Late-time accelerated de Sitter universe is also obtained when DE behaves as cosmological constant ($\Omega_{\phi}=1,~\omega_{eff}=-1$). 
These features are shown numerically in figure (\ref{uncoupled}) for parameter values $m=0.998$, and $\beta=0.001$ where the sub-fig.(\ref{noninteracting}) shows that the critical points $P_4$ and $P_5$ are the late time attractors and the point $P_1$ represents dust dominated decelerated transient evolution of the universe. The sub-fig.(\ref{noninteracting evolution}) exhibits that universe is evolving currently in quintessence era and the acceleration is driven by quintessence scalar field and the dust dominated intermediate phase is decelerating. The early scalar field dominated phase is also decelerating.

Two scaling solutions are exhibited by the points $P_6$ and $P_7$. In presence of anisotropic effect the points depict the future decelerated dust dominated universe.
In some parameters regions the points also describe the intermediate phase of decelerated evolution of the Universe.

Therefore, from the uncoupled model, we obtained the results which predict the overall evolution of the universe, namely, from early kinetic energy dominated decelerated evolution to finally late-time accelerated quintessence dominated or cosmological constant dominated de Sitter evolution of the Universe through the intermediate dust dominated decelerated phase of the Universe. In the fig. (\ref{uncoupled}) the above scenario is shown numerically for parameter values $m=0.998,~\beta=0.001$. Sub fig.(\ref{noninteracting}) shows that the points $P_2$, $P_3$ are early decelerated scalar field dominated solutions and points $P_4$, $P_5$ are scalar field dominated late-time accelerated solutions whereas the point $P_1$ describes the dust dominated decelerated transient phase of the universe. The sub-fig. (\ref{noninteracting evolution}) shows the evolution of the cosmological parameters which shows the cosmological evolution starting from scalar field dominated early decelerated universe to late time DE dominated late phase connecting through the dust dominated transition period of evolution. The late time accelerated evolution of the universe is attracted by cosmological constant in this non-interacting model. It is further to be mentioned that late time accelerated scaling solutions cannot be obtained from this model.\\

Next we studied the scalar field model of anisotropic universe in presence of interaction between dark sectors. This model has richer dynamics than that of non-interacting one. This interaction allows to achieve the late time scaling solution by exchanging energy densities and it helps to provide possible solution of coincidence problem. Our main results of the interacting model (\ref{autonomous Interacting singularity free}) is as following:\\

The set $C_1$ corresponds to a scaling solution and exists for all parameter values $\alpha,~\beta,~m$ in the phase plane. Ratio of DE-DM density parameters is $r=\frac{\Omega_{\phi}}{\Omega_{M}}=\frac{y_c^2}{1-y_c^2}$. In the range of its existence, it exhibits some physically interesting scenario. For $\alpha>0$, $y_c=0$ the set will become actually a point $(0,0)$ which will represent dust dominated future decelerated universe. Main content of the universe for this case is cosmological constant and it behaves as DM.
Physical parameters for this case are: ($\Omega_{M}=1,~\Omega_{\phi}=0,~ \omega_{\phi}=-1,~ \omega_{eff}=0$). On the other hand, for $\alpha<0$ the point (0,0) beomes a dust dominated intermediate decelerating phase of the universe and this scenario corelates with transient nature of evolution. Further, it is worthy to note that the set depicts physically relevant solution for $y_c\longrightarrow 1$ and $\alpha>0$ at late-times which favoured by present observations. The set represents the accelerated scaling attractor giving the solution of coicidence problem. 
Fig. (\ref{coupled C1 scaling}) for parameter values $m=0.9986,~\alpha=0.5,~\beta=0.5$ describes the late time scaling attractor $C_1$. In a particular value of $y_c=0.84$ (see sub-fig.(\ref{scalingC1})), we obtain a scenario supported by the present observations. Here, $\Omega_{\phi}\approx0.71,~\Omega_{M}\approx 0.29,~ \Omega_\sigma\approx 0.0028, ~ \omega_{\phi}= -1  ,~\omega_{eff}\approx -0.71,~q\approx-0.56 $. The evolution of the physical parameters show in sub-fig.(\ref{evolutionC1}) that the late-time accelerated scaling solution $C_1$ attracted in quintessence era.
\\

Completely scalar field dominated late-time acceleration of the universe is obtained by the critical points $C_6$ and $C_7$. Depending on parameter restrictions, the DE behaves as quintessence and the Universe is attracted in quintessence era only. Fig. (\ref{C6C7stable}) for $m=0.999,~\alpha=-1.8,~\beta=1$ shows that the critical points $C_6$ and $C_7$ are the scalar field dominated stable solutions. Further, the points exhibit evolution of the universe dominated by cosmological constant for $\beta=0$. The accelerated de Sitter expansion is achieved here, but not the future attractors ($\Omega_{\phi}=1,~ \omega_{\phi}=-1,~ \omega_{eff}=-1,~ q=-1$ see table \ref{physical_parameters int}).\\


Late time accelerated scaling attractors are found with the critical points $C_8$ and $C_9$. The future acceleration of universe is observed in quintessence era only and this acceleration is driven by quintessence fluid. It is worthy to note that for a particular choice of parameters, the late time attractors can alleviate the coincidence problem by satisfying $\frac{\Omega_{\phi}}{\Omega_{m}}\approx o (1)$. Also this scenario is supported by observations. Fig.(\ref{coupled C8 C9 scaling}) for $m=0.999,~\alpha=1.8,~\beta=1.999$ shows scaling attractors $C_8,~C_9$ and evolution of parameters. In particular, sub-fig.(\ref{scalingC8}) exhibits $C_8$ and sub-fig.(\ref{scalingC9}) shows $C_9$ are scaling attractors respectively. Also, in the sub-fig.(\ref{evolutionC8C9}) the late time accelerated evolution is observed in quintessence era. For $\alpha=3$, the de Sitter expansion is accelerated ($\Omega_{\phi}=1,~ \omega_{eff}=q=-1$) but the de Sitter solution cannot be future attractor.  Depending upon some parameter restrictions, the points also can be accelerated and past attractors in phase plane. This scenario corelates with the early inflationary evolution of the universe. For example, see the figure (\ref{coupled complete evolution}) for parameter values $m=0.95,~\alpha=2.9,~\beta=2.23297$ where a complete cosmic scenario of evolution is achieved from early inflation to a late time accelerated DE dominated evolution passing through a DM dominated decelerating transient phase of the universe. Here, critical points $C_8$ and $C_9$ describe the early inflation having their corresponding physical parameters as $\Omega_{M}\approx 0.03,~\Omega_{\phi}\approx 0.97,~\omega_{\phi} \approx -0.99,~\omega_{eff}\approx -0.97,~q\approx -0.97$,  and the corresponding eigenvalues are positive ($\lambda_1>0,~\lambda_2>0$) showing that the points are source in phase plane and this scenario describes the early phase of evolution. In this figure, points $C_6$ and $C_7$ describe an intermediate phase of evolution. Although, the points are completely DE dominated ($\Omega_{\phi}=1$), other parameters indicate that the points mimic the dust dominated transient phase of the universe ($\omega_{eff}\approx 0,~q\approx 0.5$). Here, the DE mimics as dust ($\omega_\phi\approx 0$).
Finally, the set $C_1$ exhibits the late time scaling attractor in the phase plane. For a particular choice of $y_c$, the point with coordinates $(0,~0.84)$ and ($0,~-0.84$) on the set represents late time DE dominated accelerated evolution of the universe where the corresponding cosmological parameters are $\Omega_{\phi}=0.71,~\Omega_{M}=0.29,~\omega_{\phi}=-1,~\omega_{eff}=-0.71$.
Therefore, in conclusion one can draw a sequence of critical points $C_8 (\mbox{inflation})\longrightarrow C_6 (\mbox{matter dominated transition})\longrightarrow  C_1 (\mbox{late phase})$. This is shown in sub-fig.(\ref{complete scenario2}) where trajectories are coming out of $C_8$, entering to the point $C_6$ and emerging from it to go finally to the point $C_1$. The similar is shown in sub-fig.(\ref{complete scenario3}) where the trajectories starting from the point $C_9$, entering to the point $C_7$ and finally going to the point $C_1$. That is $C_9(\mbox{inflation}) \longrightarrow C_7(\mbox{matter dominated transition}) \longrightarrow C_1(\mbox{late phase})$ showing the complete evolutionary scheme of the universe. The evolution of $\omega_{eff},~q$ and $\omega_\phi$ (see in sub-fig.(\ref{complete scenario evolution})) shows that the cosmological evolution of the universe is started where $\omega_{eff}\approx-1$ describes the early inflation and acceleration is driven by scalar field with $\omega_\phi\approx -1$, passing through a phase(called intermediate phase) where $\omega_{eff}\approx 0$ and $\omega_\phi\approx 0$ (i.e., scalar field DE behaves as dust), and finally the evolution of the universe is attracted in a regime where $\omega_{eff}<-\frac{1}{3}$ and $\omega_\phi\approx -1$ (i.e., late time acceleration). This is the most relevant and observationally fit model of the universe which is obtained from our study.
\\

Therefore, one can conclude that the model for non-interacting and interacting DE with pressureless dust providing two autonomous systems namely, Eqns (\ref{autonomous noninteracting}) and (\ref{autonomous Interacting singularity free}) yields several physically interested solutions. First,  dust dominated intermediate decelerating phase is achieved for non-interacting as well as for interacting one. Secondly, scalar field dominated accelerated phase of evolution is also obtained for the two cases as well. Scaling solutions are found in both the cases, but only in interacting case, the scenario is cosmologically interesting. Because, scaling solution is accelerated late time attractor providing the mechanism to alleviate the coincidence problem. On the other hand, in non-interacting case, the scaling solutions can give only decelerated dust solution either in future evolution or in intermediate phase due to some parameter restrictions. Finally, a complete cosmic scenario starting from early inflation to late phase of universe, conneting through a matter phase can be achieved only for interacting scenario. 

\section{Comparison of the Theoretical Model with the Observational
Data}\label{cdata}
From the energy conservation relations for the subsystems given by (\ref{conservation DM}) and (\ref{conservation DE}), the energy densities of DM and DE are calculated as follows:

\begin{equation}\label{equation28}
\rho_M= \rho_{M0} \left(AB^{2}\right)^{\frac{\alpha-3}{3}} 
\end{equation}
and
\begin{equation}\label{equation29}
\rho_\phi= \rho_{\phi 0} \left(AB^{2}\right)^{-(1+\omega_{\phi})}+\rho_{M0} \frac{\alpha}{\alpha+3\omega_{\phi}}\left[ (AB^2)^{-(1+\omega_{\phi})} -(AB^2)^{\frac{\alpha-3}{3}} \right]
\end{equation}
where $\rho_{M0}$, $\rho_{\phi 0}$ are present value of energy densities of DM and DE respectively and $\omega_\phi=\frac{p_\phi}{\rho_\phi}$ is equation of state parameter (EoS) of DE.
Then using the above expressions for energy densities and equation (\ref{Friedmann}) we obtain analytic expression for Hubble parameter $H(z)$ as
\begin{equation}\label{equation30} \centering \fontsize{9.5pt}{5pt}
H=H_{0}\left[ \Omega_{M0}\left( \frac{1}{1+z}\right) ^{\left( 1+\frac{2}{n})(\frac{\alpha-3}{3}\right) }+\Omega_{\phi0}\left( \frac{1}{1+z}\right) ^{-\left( 1+\frac{2}{n}\right)(1+\omega_{\phi}) }   + \frac{\Omega_{M0} \alpha}{\alpha+3\omega_{\phi}} \left(  (\frac{1}{1+z})^{-(1+\omega_{\phi})(1+\frac{2}{n} )} -   (\frac{1}{1+z})^{(1+\frac{2}{n})(\frac{\alpha-3}{3})}  \right)\right]^{\frac{1}{2}} 
\end{equation}
where $z=\frac{1}{A}-1=\frac{1}{B^{n}}-1$ is redshift parameter.\\

\begin{figure}
\centering
\subfigure[]{
\includegraphics[width=7cm,height=6cm]{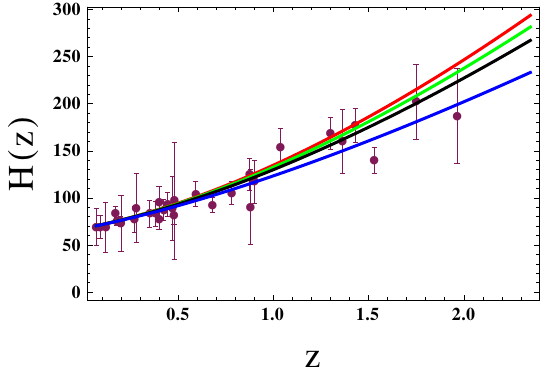}\label{hhh}}
\qquad
\subfigure[]{
\includegraphics[width=7cm,height=6cm]{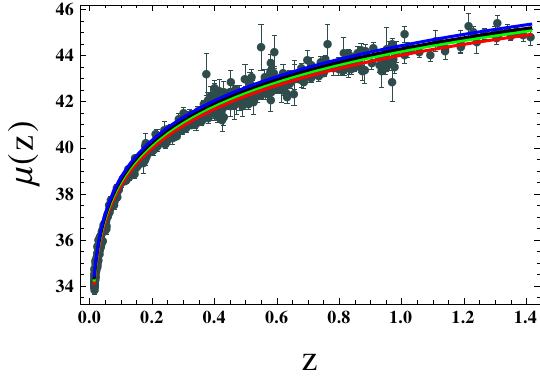}\label{mu}}
\qquad
\caption{Figures (a) and (b) represent the comparison of our obtained model with $1\sigma$ error bars of 31 data points for $H (z)$ \cite{hdata1,hdata2,hdata3} and 580 data points for Supernova Type Ia (Union 2.1 compilation dataset) \cite{580snia} observations, respectively.  In panel (b), $\mu(z)=5{\rm log_{10}}(d_{L}/Mpc)+ 25$ (with $d_{L}$ being the luminosity distance) \cite{cst2006}, denotes the distance modulus, which is the difference between the apparent magnitude and the absolute magnitude of the observed Supernova. In each panel, the red, green, black and blue lines are for $\alpha=0$ (non-interacting case), $0.09, 0.2$ $\&$ $0.5$, respectively. All the plots are for $\Omega_{M0} = 0.315$, $H_{0}=67.4$ km/s/Mpc \cite{hub674}, $n=0.85$ (or equivalently $m=0.9986$) and $\omega_{\phi}=-0.95$. } 
\label{dobs1}
\end{figure}

In figure \ref{hhh}-\ref{mu}, we have shown the evolutions of $H(z)$ vs. $z$  and  $\mu(z)$ vs. $z$ for our  model and compared it with that of the 31 data points for $H(z)$ and 580 data points for Supernova Type Ia measurements  with $1\sigma$ error bars, respectively. It has been found that for most of the cases, the considered model is consistent with the $H(z)$ and $\mu(z)$ datasets at low redshifts ($z<1$) for different values of the parameter $\alpha$. Moreover, we have checked that the evolutionary behaviors of $H(z)$ and $\mu(z)$  are hardly affected by a small change in the values of the model parameters. 

\clearpage
\clearpage
\section{Concluding remarks}\label{sec-last}

We investigated the cosmological dynamics of a scalar field model minimally coupled to gravity in a spatially flat, homogeneous, Locally Rotationally Symmetric (LRS) Bianchi type I anisotropic Universe where the scalar field taken as DE and pressureless dust as DM is the main content. We considered a pre-factor, termed a coupling function $\lambda(\phi)$, in kinetic terms of the scalar field. The Lagrangian, in presence of coupling, is defined as the action in (\ref{action}). We analyzed uncoupled and coupled models separately with the help of dynamical systems tools and techniques. For coupled model, we considered the interaction term (\ref{interaction}) from a phenomenological point of view and for mathematical simplicity.   
First, we converted the cosmological evolution equations into an autonomous system of ODEs by a suitable transformation of variables that are dimensionless and normalized over the Hubble scale. 
The nature (stability) of critical points is found by evaluating the eigenvalues of the Jacobian matrix at the critical points. Linear stability theory is employed for characterizing the points.
The coupling function $\lambda(\phi)$ and the potential $V(\phi)$ of the scalar field are chosen in such a way that the term $\beta(\phi)=\sqrt{\frac{3}{2}} \frac{V^{'}(\phi)}{V(\phi) \sqrt{\lambda(\phi)}}$, appeared in the system of ODEs in (\ref{non-autonomous}), is to be a constant. As a result, we obtained a two-dimensional autonomous system to study.  We showed the anisotropic effects on the stability of each of the critical points in detail. The anisotropic density is measured by $\Omega_\sigma=1-m^2$, where $-1\leq m<0,~~0<m\leq 1$. For $m=\pm 1$, the anisotropic effect vanishes, and the usual FLRW model is recovered. 
\\

We studied first, a non-interacting model of a 2D autonomous system by virtue of linear stability analysis in subsection \ref{non-interacting}. 
The critical points extracted from the autonomous system are all hyperbolic types. The critical points and the corresponding physical parameters are presented in table \ref{physical_parameters}.

The critical point $P_1$ always exists and represents a saddle-like solution in the phase plane. It depicts the dust-dominated intermediate decelerating era of the evolution (see in fig. (\ref{uncoupled})).
Critical points $P_2$ and $P_3$ are potential of scalar field dominated solutions. Accelerating universe is not possible near these points. The points are not of interest in late-time cosmology. However, depending on the parameters $(\beta, ~m)$ the points may be past attractors in the phase plane.
Scalar field dominated solutions $P_4$ and $P_5$ are the same in all respect. The points exhibit the late-time accelerated evolution of the universe in a quintessence era where DE also behaves as quintessence due to some restrictions on $\beta$ and $m$ (see in subsection \ref{non-interacting}). Moreover, the points also correspond to accelerated de Sitter expansion at late times where acceleration is driven by cosmological constant fluid and the universe is dominated by the potential of scalar field.
Two scaling solutions, $P_6$ and $P_7$, are not physically attractive at late times because they are constantly decelerating in nature. However, they can predict the intermediate decelerated evolution of the universe when the points are saddle-like solutions in the phase plane, and this is possible for parameter conditions. Also depending parameters $\beta$ and $m$ the points can be late time solutions. In this case, we can conclude that the future deceleration of the universe is predicted only by points $P_6$ and $P_7$.\\

Next, we studied the above model by considering an interaction term in the dark sectors. We obtained the autonomous system (\ref{autonomous Interacting singularity free}), where an extra parameter $\alpha$, called the coupling parameter of the interaction, appeared in the system. As a result, the analysis is now dependent on three parameters $\alpha$, $\beta$ and $m$. For $\alpha=0$, the system reduces to that of uncoupled one in (\ref{autonomous noninteracting}). However, we have obtained some critical points showing exciting features from a cosmological point of view. 

First of all, a normally hyperbolic set  $C_1$  exists for $0\leq y_c^2\leq1$ in the phase plane. It is a solution with a combination of both the DE and DM. The set represents DM-dominated decelerating evolution of the universe for $y_c=0$  ($\Omega_{M}$=1, ~$\omega_{eff}$=0, q=$\frac{1}{2}$), interestingly, cosmological constant behaves here as DM. The set exhibits DE-dominated late-time acceleration of the universe when $y_c\longrightarrow 1$.
The critical points $C_2,~C_3$ are the same as the points $P_2,~P_3$ in the non-interacting model when $\alpha=0$. They are scalar field-dominated decelerated solutions, not of much interest in late times.
The critical points $C_4$ and $C_5$ are new in the interacting model. The points exist only in $\alpha \in(-3,~0)$. The points are scaling solutions in the phase plane. Depending on the parameters $(\alpha, \beta, m)$ values,  the points may represent stable solutions, but they are not physically interested because the points are constantly decelerating in their existence region. However, they can predict the future deceleration of the universe through these points.
Points $C_6$ and $C_7$ are similar to that of points $P_4$ and $P_5$ in non-interacting case. Only stability is differed due to the appearance of the coupling parameter $\alpha$. The points are completely dominated by scalar field. Late time accelerated evolution is predicted through these points where the universe evolves in quintessence era, and DE behaves as quintessence. 
Late-time accelerated scaling attractors are found in this model at the points $C_8$ and $C_9$. Depending on parameter values, these points show various interesting scenarios in the phase plane. For $\alpha\longrightarrow 3$, the points are dominated completely by scalar field and upon restricted values of other parameters $\beta$ and $m$, accelerated past attractors can be achieved, similar to the early inflationary scenarios.

On the other hand, the points can predict future attractors satisfying the similar order of energy densities of matter and scalar field which leads to help in solving the coincidence problem. This solutions cannot be obtained in the non-interacting scenario.\\ 

Therefore, depending on model parameters ($\alpha, \beta$) and the anisotropy effect (measured by $m$) on space-time, the model provides some physically attractive solutions in late-times as well as at early times. The main findings are: the scalar field-dominated solutions characterized by points $C_6$, $C_7$ corresponding to the accelerating universe in the quintessence era at late-times. However, they need to solve the coincidence problem. Scalar field-dark matter accelerated scaling solutions (characterized by set $C_1$ and the points $C_8,~C_9$) mimic late-time attractors acquiring a similar order of density parameters for DE and DM. This alleviates the coincidence problem and is supported by present observations also. The corresponding figures are plotted in the previous section. Finally, a unified cosmic evolution is obtained from early inflation to DE-dominated late-times acceleration connecting through a matter-dominated transient phase. It is shown in fig. (\ref{coupled complete evolution}) that the evolution occurred as $C_8 (\mbox{inflation})\longrightarrow C_6 (\mbox{matter- dominated transition})\longrightarrow  C_1 (\mbox{late phase})$  or $C_9(\mbox{inflation}) \longrightarrow C_7(\mbox{matter-dominated transition}) \longrightarrow C_1(\mbox{late phase})$. Particularly, numerical investigation shows in sub-fig.(\ref{complete scenario evolution}) that a cosmological evolution where $\omega_{eff}$ evolves starting from an early inflationary era ($\omega_{eff}\approx -1$), passing a matter phase ($\omega_{eff}\approx 0$) and finally being attracted late -time dark energy phase ($\omega_{eff}<-\frac{1}{3}$).  \\
\par Lastly, we investigated the evolutions of the Hubble parameter and the distance modulus for the non-canonical scalar field model under consideration and
compare that with the observational Hubble parameter and Supernova Type Ia datasets (see Figures \ref{hhh} $\&$ \ref{mu}). For different values of $\alpha$, we found that the trajectories of the Hubble parameter and the distance modulus predicted by the model are in quite good agreement with those observational datasets. Interestingly, dynamical study of this interacting model also shows the same results. For a similar data set ($m=0.9986$ (or equivalently $n=0.85$), $\alpha=0.5$, $\beta=0.5$), the fig.(\ref{coupled C1 scaling}) exhibits the evolution of universe supported by present observations.\\

\par In summary, we now conclude that our model gives some essential and exciting consequences from the cosmological perspective. Additionally, it would be interesting to study the effect of perturbation on the non-canonical scalar field model in the framework of anisotropic LRS Bianchi type I universe. However, this study is left for the future work.\\
\section*{Acknowledgments}
The authors would like to thank the anonymous referee for constructive comments. Sujoy Bhanja is grateful to UGC (University Grants Commission), Govt. of India for giving Junior Research Fellowship [F. No: 16-6(DEC.2018)/2019(NET/CSIR) UGC-Ref. No.:1140/(CSIR-UGC NET DEC. 2018)] for the Ph.D work. Goutam Mandal is supported by UGC, Govt. of India through Senior Research Fellowship [Award Letter No. F.82-1/2018(SA-III)] for Ph.D.


\newpage

\end{document}